\documentclass[times,twocolumn,final]{elsarticle}

\usepackage{jasr}
\usepackage{framed,multirow}
\usepackage{latexsym}
\usepackage[switch]{lineno}
\usepackage{url}
\usepackage{xcolor}
\definecolor{newcolor}{rgb}{.8,.349,.1}

\usepackage[citebordercolor=white]{hyperref}
\usepackage{natbib}
\usepackage{mathptmx}
\usepackage{helvet}
\usepackage{courier}

\usepackage[T1]{fontenc}
\usepackage[latin9]{inputenc}
\usepackage{color}
\usepackage{babel}
\usepackage{amsbsy}
\usepackage{amstext}
\usepackage{amssymb}
\usepackage{graphicx}
\usepackage{esint}

\makeatletter

\providecommand{\tabularnewline}{\\}

\definecolor{newcolor}{rgb}{.8,.349,.1}\usepackage{babel}

\usepackage{color}
\usepackage{babel}
\usepackage{amstext}

\defcitealias{MalkMosk2022}{Paper~II}
\defcitealias{MalkMosk_2021}{Paper~I}

\def\la{\mathrel{\mathchoice {\vcenter{\offinterlineskip\halign{\hfil
\(\displaystyle##\)\hfil\cr<\cr\sim\cr}}}
{\vcenter{\offinterlineskip\halign{\hfil\(\textstyle##\)\hfil\cr
<\cr\sim\cr}}}
{\vcenter{\offinterlineskip\halign{\hfil\(\scriptstyle##\)\hfil\cr
<\cr\sim\cr}}}
{\vcenter{\offinterlineskip\halign{\hfil\(\scriptscriptstyle##\)\hfil\cr
<\cr\sim\cr}}}}}

\makeatother

\journal{Advances in Space Research}

\begin{document}
\begin{frontmatter}
\title{Very Local Impact on the Spectrum of Cosmic-Ray Nuclei below 100 TeV}

\author[1]{M.A. \snm{Malkov}\corref{cor1}}
\cortext[cor1]{Corresponding author: mmalkov@ucsd.edu}
\author[2]{I.V. \snm{Moskalenko}}
\author[1]{P.H. \snm{Diamond}}
\author[1]{M. \snm{Cao}}


\affiliation[1]{organization={Department of Astronomy and Astrophysics, University of California San Diego},
                city={La Jolla},
                postcode={CA 92093},
                country={USA}}

\affiliation[2]{organization={Hansen Experimental Physics Laboratory and Kavli Institute
for Particle Astrophysics and Cosmology, Stanford University},
                city={Stanford},
                postcode={CA 94305},
                country={USA}}

\received{\today}
\finalform{}
\accepted{}
\availableonline{}

\begin{abstract}
Recent measurements of primary and secondary CR spectra, their arrival
directions, and our improved knowledge of the magnetic field geometry
around the heliosphere allow us to set a bound on the distance beyond
which a puzzling 10-TeV ``bump'' and certain related spectral features
\emph{cannot} originate. The sharpness of the spectral breaks associated
with the bump, the abrupt change of the CR intensity across the local
magnetic equator ($90^{\circ}$ pitch angle), and the similarity between
the primary and secondary CR spectral patterns point to a local reacceleration
of the bump particles out of the background CRs. We argue that, owing
to a steep preexisting CR spectrum, a nearby shock may generate such
a bump by boosting particle rigidity by a mere factor of $\sim$1.5
in the range below 50 TV. Reaccelerated particles below $\sim$0.5
TV are convected with the interstellar medium flow and do not reach
the Sun. The particles above this rigidity then form the bump. This
single universal process is responsible for the observed spectral
features of all CR nuclei, primary and secondary, in the rigidity
range below 100 TV. We propose that one viable candidate is the system
of shocks associated with $\epsilon$ Eridani star at 3.2 pc of the
Sun, which is well aligned with the direction of the local magnetic
field. Other shocks, such as old supernova shells, may produce a similar
effect. We provide a simple formula that reproduces the spectra of
all CR species with only three parameters uniquely derived from the
CR proton data. We show how our formalism predicts helium, boron,
carbon, oxygen, and iron spectra. Our model thus unifies all the CR
spectral features observed below 50 TV. 
\end{abstract}
\end{frontmatter}

\section{Introduction}

Sources of cosmic rays (\textbf{CR}s) are scattered across the \emph{Galaxy.}
After propagation to the Earth, accompanied by a turbulent reacceleration
in stochastic magnetic fields and a partial leak into intergalactic
space, CRs are expected to arrive with a featureless power-law energy
spectrum. The angular distribution should be weakly anisotropic with
a dipolar component $\sim$$10^{-3}$,  indicating that more particles
come from the \emph{inner} Galaxy (Secs.\ref{subsec:Anisotropy} and
\ref{subsec:Anisotropy2}). However, the new observations revealed
a prominent structure in the \emph{spectra of CR primaries} (so-called
10-TeV ``bump'') also characterized by \emph{an excess pointing
to the outer} Galaxy with a sharp step-like boundary in CR arrival
directions and similar spectral anomalies in secondary CR \emph{species}.

Understanding the origin of the bump offers a deeper insight into
the nature of CR propagation from their sources, individual source
contributions to the CR spectrum, and the structure of the local \emph{interstellar
medium (ISM).} Upon addressing the new challenges, we can test and
improve the concept of CR origin and guide future observations.

This paper focuses on CR \emph{reacceleration and propagation} \emph{in
the local} ISM ($\sim$10 pc of the Sun). An exceptional sharpness
of the breaks in the \emph{spectra of CR species} resolved by AMS-02,
CALET, DAMPE, and ISS-CREAM orbital instruments, dictates this choice.
\emph{These features persist within unprecedentedly tight error bounds.}
Local SNRs at the distances of 100s pc are unlikely responsible for
these structures. Propagating CRs from \emph{such sources} would erase
fine spectral features, such as sharp breaks in rigidity spectra and
small-scale angular anisotropies.

On the other hand, a nearby passing star, blowing its wind against
the headwind of the ISM, can make an efficient CR \emph{reaccelerator}
within 3-10 pc of the Sun. A synergy of the star's bow- and wind termination
shocks, \emph{which can be supplemented by reacceleration in the stellar
wind, is able to} efficiently reaccelerate the preexisting CRs. Propagating
then through a turbulent magnetic flux tube pointed to the Sun, the
reaccelerated CRs most naturally acquire the observed spectral features
by suffering energy-dependent diffusive and convective losses from
the tube.

\begin{figure*}[tb!]
\centering
	\includegraphics[scale=0.18]{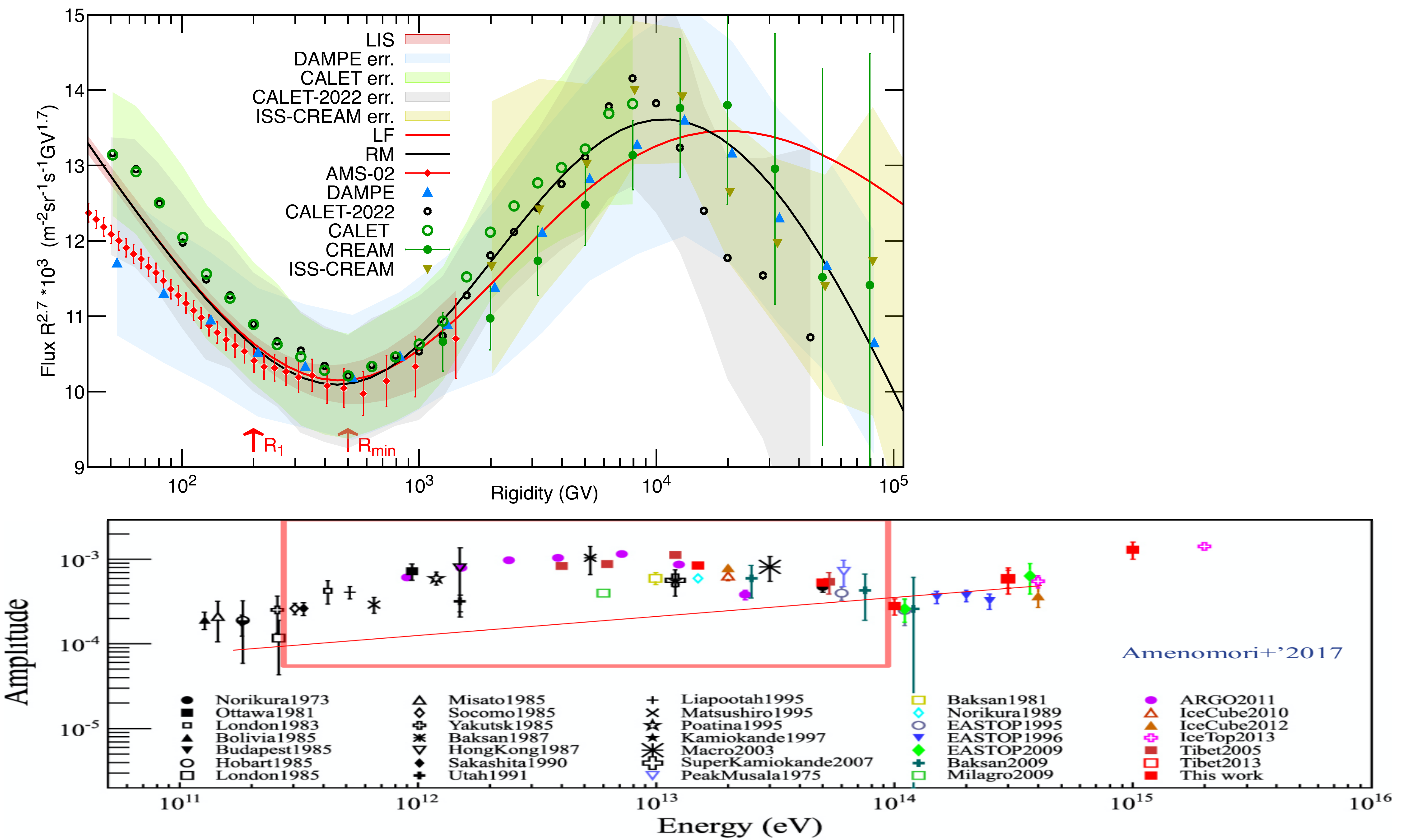} \caption{\textbf{Top: }New prominent spectral features, shown in the $10^{2}$--$10^{5}$
		GV range for protons. Compilation of data from AMS-02 \citep{Aguilar2021},
		CALET \citep{CALET_2019PhRvL,AdrianiCalet2022}, CREAM \citep{2017ApJ...839....5Y},
		ISS-CREAM \citep{ISS-Cream2022}, and DAMPE \citep{Dampe2019a} is
		described in Sec.\ \ref{sec:Observational-Challenges}. An arrow labeled $R_{1}$ shows the break rigidity. Note that
		the shown CALET, DAMPE, and ISS-CREAM data are adjusted to match the
		ISM spectrum \citep{2020ApJS..250...27B} at the 
		minimum at $R_{min}\approx$500 GV. This requires shifts in both the rigidity
		and normalization, which are individual for each set of data. In particular,
		the older CALET \citep{CALET_2019PhRvL} and DAMPE data points are
		shifted as described in \citet{MalkMosk2022}. The adjustments used
		for the older CALET data are also applied to the newer CALET data
		set \citep{AdrianiCalet2022}. The CREAM and ISS-CREAM data points
		are renormalized by a factor of 1.12 to match the CALET data \citep{AdrianiCalet2022}.
		The local interstellar spectrum (LIS) is taken from \citet{2020ApJS..250...27B},
		where the marked uncertainty corresponds to the uncertainty in the
		AMS-02 data. The lines show the local \emph{ISM} spectra in two models,
		Eq.~(\ref{eq:SolFinal}): \emph{Red --} without CR losses from the
		magnetic flux tube, \emph{Black --} the same fit with losses (Tables~\ref{tab:Model-parameters-and}
		and \ref{tab:Input-parameters-and}).\textbf{ Bottom:} CR anisotropy
		data with the energy scale aligned with the top panel, adopted from \citet{Amenomori2017}.}
	\label{fig:Bump}
\end{figure*}
\section{Observational Challenges\protect\label{sec:Observational-Challenges}}

\subsection{Energy Spectrum}

The high-precision data accumulated by several new instruments, AMS-02
\citep{Aguilar2021}, CALET \citep{AdrianiCalet2022}, DAMPE \citep{Dampe2019a},
ISS-CREAM, \citep{ISS-Cream2022}, challenge the CR acceleration and
propagation theories. Fig.\ \ref{fig:Bump} shows a zoomed $10$-TeV
bump in an enhanced rigidity format with an extra $R^{2.7}$ factor.
The data have revealed two prominent features. The deviations from
the straight power law begin with a sudden flattening at $\approx$$0.2$
TV. Then, the spectrum steepens back at $\approx$$10$ TV, also very
sharply. It might then return to its low-energy slope in the 100 TV
range. This spectral anomaly is a challenge since the acceleration
of CRs in SNR shocks and their subsequent propagation to the heliosphere
was long considered scale-invariant.

\begin{figure}
	\includegraphics[scale=0.24]{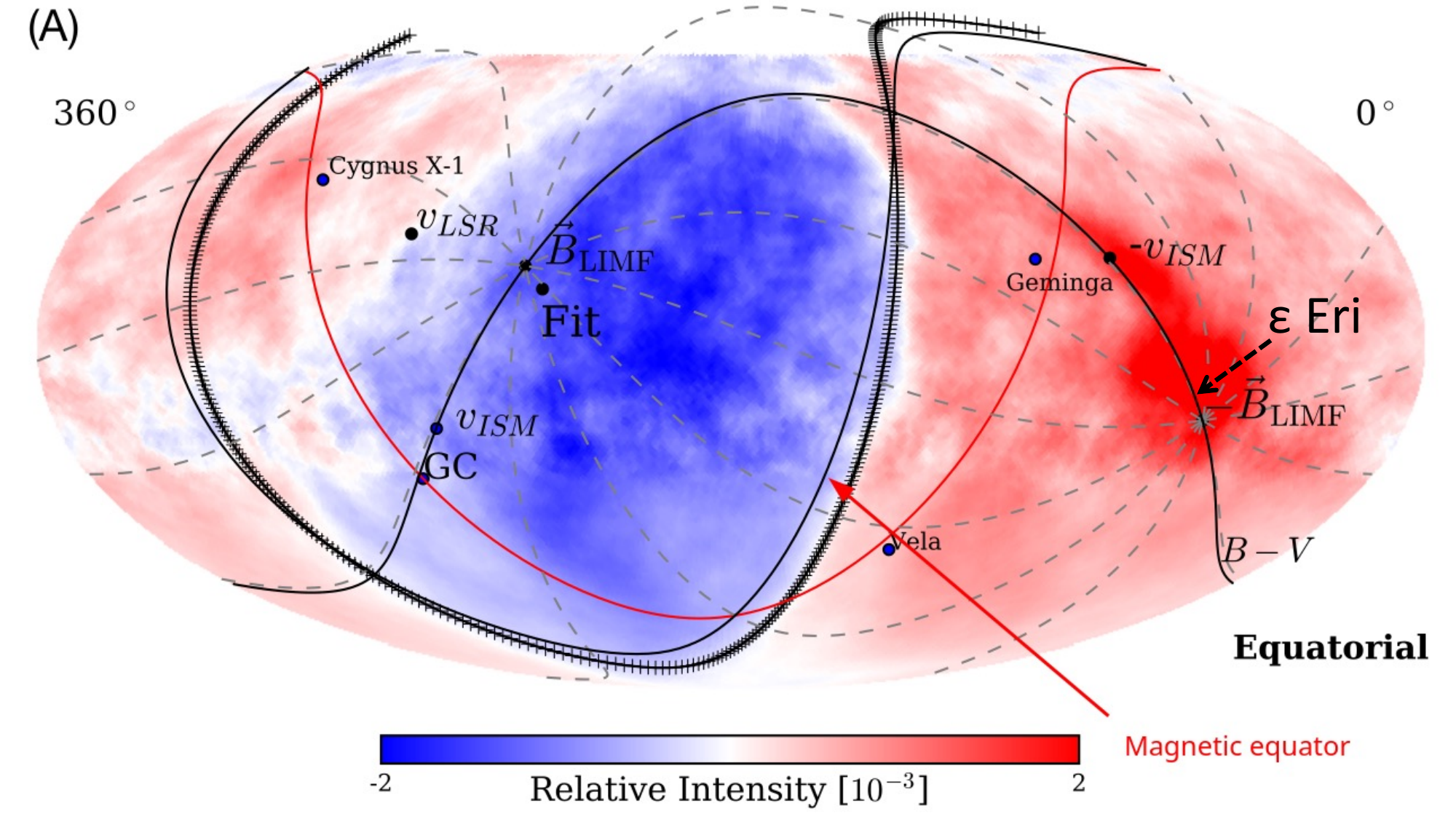} \caption{CR arrival direction intensity \emph{at median primary particle energy
			of 10 TeV} adopted from \citet{Abeysekara2019} (Fig.11a, with added
		arrows pointing to the magnetic equator and $\epsilon$ Eridani position.)}
	\label{fig:Anisotropy} 	
\end{figure}

An arrow labeled $R_{1}$ in Fig.\ \ref{fig:Bump} shows the break rigidity. Note that
the shown CALET, DAMPE, and ISS-CREAM data are adjusted to match the
ISM spectrum \citep{2020ApJS..250...27B} at the 
minimum at $R_{min}\approx$500 GV\footnote{Although $R_{\text{min}}$ appears to be a convenient visual characterization
of the break position, it depends on the spectrum enhancement factor.
The quantity $R_{1}$ is instead characterized by a statistically
significant deviation of the data points from a power-law below the
break.}. This adjustment requires shifts in both the rigidity and normalization,
which are individual for each set of data. It is not critical for
our model  that relies only on the shape of the CR proton spectrum.
In particular, the older CALET \citep{CALET_2019PhRvL} and DAMPE
data points are adjusted as described in \citep[Paper~II]{MalkMosk2022}. The
adjustments used for the older CALET data are also applied to the
newer CALET data set \citep{AdrianiCalet2022}. The CREAM and ISS-CREAM
data points are renormalized by a factor of 1.12 to match the CALET
data \citep{AdrianiCalet2022}.

\subsection{Anisotropy\protect\label{subsec:Anisotropy}}

The anisotropy data shown in Fig.\ \ref{fig:Bump} (bottom panel)
are widely interpreted as almost flat in a broad energy range between
$10^{11}$--$10^{15}$ eV, within uncertainties. This flatness is
surprising, as the large-scale anisotropy is expected to grow with
particle energy because they leave the Galaxy faster. Here we note,
however, a striking coincidence between the energy range of the bump
Fig.\ \ref{fig:Bump}, and the anisotropy enhancement between $\approx$$2\times10^{11}-10^{14}$
eV, marked by a box. After \emph{that}, the anisotropy briefly declines
toward $\sim$$10^{14}$ eV \emph{and then grows again.} The enhanced
anisotropy in the box area that can be primarily associated with the
10-TV rigidity bump in Fig.\ \ref{fig:Anisotropy} strongly suggests
the flat anisotropy being composed of two independent CR sources:
the background CRs with a weak but growing anisotropy, as expected,
masked by a separate component comprising the bump particles.

The overall anisotropy discussed above is at a $\sim$$10^{-3}$ level,
corresponding to a dipole component of CRs. There is also a small-scale
anisotropy at $10^{-4}$, first discovered by the Milagro observatory
\citep{Milagro08PRL}. More accurate mapping is provided by HAWC and
IceCube \citep{Abeysekara2019}. Fig.\ \ref{fig:Anisotropy}\emph{
shows a well-defined $\sim$$20^{\circ}$-wide excess likely associated
with the CR prevalent arrival direction, and may very well be relevant
to the 10-TeV CR bump in the rigidity spectrum shown in Fig.\ref{fig:Bump}.
}The position of the excess on the map shows that these particles
are moving along the local magnetic field toward the \emph{inner galaxy.
Moreover, the position of $\epsilon-$ Eridani star is at the center
of the CR excess.}  Another striking aspect of the small-scale anisotropy,
which is particularly crucial to our assertion of the proximity of
its source, is a sharp increase of the CR intensity across the magnetic
equator, as seen in Fig.\ \ref{fig:Anisotropy}. A straightforward
interpretation of this increase is that particles pass the observer
while moving from the outer to the inner galaxy. They have neither
been scattered appreciably by the magnetic turbulence nor mirrored
by it. Notably, the source must be located somewhere not far from
the observer on the outer galaxy side. We will discuss  these aspects
of angular anisotropy at a quantitative level in Sec.\ \ref{subsec:PlausibilityOfLocal}.

A review of recent observations of the CR anisotropy and theoretical
models is given in \citet{BeckerTjus2020}. The latest data covering
a relatively wide swath of the Northern and Southern hemispheres were
reported by the GRAPES-3 team \citep{Chakraborty2023}. These data
are largely consistent with the HAWC/IceCube small-scale counterpart
of the map shown in Fig.\ \ref{fig:Anisotropy}. Given
the close proximity of the object, which we suggest responsible for
the observed anisotropy, a temporal evolution of the map compared
to the earlier Milagro data is quite possible \citep[Paper~I]{MalkMosk_2021}.

\subsection{Secondaries}

Secondary species exhibit stronger spectral hardening than the primaries
\citep{Aguilar2021}. It takes millions of years to produce the secondaries
out of the primaries in CR spallation reactions. Matching the observed
breaks in secondaries with additional components produced in local
sources surrounded by gas clouds requires fine-tuning.


\section{How Far Away the TeV-bump May Have Formed?\protect\label{sec:Solar-Neighborhood-Origin}}

Whatever the reasons for the breaks, they should form not too far
from us to survive smoothing by the diffusion in momentum. Using a
simple relation \citep{Skill75a} between the momentum diffusivity,
$D_{p}$, and the spatial diffusivity, $\kappa$, $D_{p}\approx p^{2}V_{A}^{2}/\kappa$,
where $V_{A}$ is the Alfv\`en velocity, one can estimate the distance
to the source as follows.

A sharp break spreads in momentum by $\Delta p/p\sim\sqrt{D_{p}t_{\text{prop}}}/p\sim LV_{A}/\kappa$,
where $L\sim\sqrt{\kappa t_{\text{prop}}}$ the distance to its origin
and $t_{\text{prop}}$ is the propagation time. The width $\Delta p$
remains small for $L\ll\kappa/V_{A}$. For energies above a few GeV,
$\kappa\sim10^{28}\left(p/[\text{GeV}/c]\right)^{\delta}\text{cm}^{2}/\text{s}$,
$\delta=0.3-0.6$, and $V_{A}\simeq30$ km/s, we arrive at the following
restriction $L_{s}\ll1$ kpc. This restriction is rigid if the first
break is as sharp as the most precise AMS-02 data suggest. More importantly,
since the number density of CRs at the bump is \emph{a factor of}
$\sim$2.4 \emph{higher than} the background CRs underneath it, these
particles likely drive Alfv\`en waves by themselves. Simple estimates
suggest a decrease in $\kappa$ by at least an order of magnitude
\citepalias{MalkMosk_2021}. This estimate strictly limits the distance
to the source to $L_{s}\ll10^{2}$ pc. The above estimate concerned
only the propagation history of the bump. Its formation in a nearby
SNR, as often suggested, must be accompanied by an even stronger spread
in momentum because of an even more substantial reduction in $\kappa$
around the source \citep{MetalEsc13}, resulting in a delayed particle
escape from its surroundings. We conclude that the bump must have
formed not farther than a few tens of parsecs. However, the bump is
unlikely to form inside or near the heliosphere.

As one of the angular CR enhancements discovered by Milagro \citep{Milagro08PRL}
was in the heliotail direction, attempts were made to explain it by
the acceleration of CRs in the regions of a striped magnetic field
in the heliotail. Other peripheral magnetic perturbations around the
heliosphere have also been examined. However, the 10 TeV proton gyroradius,
$r_{g}$, in the 1-3 $\mu$G field is $\sim$$10^{3}$ au, an order
of magnitude larger than the heliosphere cross-section radius in its
tail region, $R_{\text{hs}}$. CR acceleration mechanisms, based,
e.g., on the reconnection in the striped magnetic field in the tail,
would require $R_{\text{hs}}\gg r_{g}$, while an $R_{\text{hs}}\lesssim r_{g}$
condition is guaranteed for 10 TeV protons.

Secondary species provide another compelling argument against the
nearby SNR origin of the 10-TeV bump. They exhibit a stronger spectral
hardening than the primaries at the same rigidity. It indicates a
rigidity-dependent modification of the spectra, possibly associated
with the diffusion or diffusive reacceleration rather than the energy-per-nucleon
spallation-related process, given an insufficient material grammage
on the path to the observer from a nearby SNR.

The above two lines of argument imply seemingly opposite restrictions
on the distance to the source of the spectral bump. On the one hand,
these particles must spend a very long time in the Galaxy, thus propagating
a very long distance, to process a significant ``grammage'' of the
ISM gas. On the other hand, they somehow evade the momentum diffusion,
which is inevitable in the turbulent ISM. The sharp spectral breaks
in particle momentum distribution point to a relatively local origin
of the breaks, while the underlying particles must have been accelerated
in distant objects, most likely an ensemble of SNRs, to process sufficient
grammage. The most natural way to reconcile this contradiction is
to assume that the bump comprises \emph{locally reaccelerated }background
CRs.

Alternatively, matching the Galaxy-wide spectra with the local-source-produced
component requires fine-tuning the local source abundances, its surrounding
gas grammage, and the rigidity dependence of the diffusion coefficient
in the whole Galaxy \citep[e.g.,][]{2023ApJ...956...75Q}. Apart from the fine-tuning and 
multi-source invocation, it is not clear whether the respective models fit the high-fidelity data shown, e.g.,  in Fig.\ref{fig:Bump} (top panel). These models typically span 2-4 orders of magnitude in the CR flux (multiplied by $R^{2.7}$), as opposed to the factor $\simeq$$1.5$ shown in the figure. As can be gleaned from Fig.\ref{fig:Anisotropy}, the presumed magnetic connectivity of the suggested local sources, such as Geminga,\citep[e.g.,][]{Zhao_2022}, is questionable for at least two reasons. First, its position on the CR map is not well aligned with either the local field direction or the CR hot spot, which is in striking contrast with the $\epsilon$ Eridani. Second, the field direction at the sun is almost certainly irrelevant to that at any source outside of the Local Bubble, such as Geminga SNR. This lack of magnetic connectivity owes to the diamagnetic flux expulsion during the Local Bubble formation by a series of SN explosions some millions of years ago. We will discuss other models of the 10-TV bump origin outside of the Local Bubble in Sec.\ref{sec:Alternative-Explanations-of}.

\section{Plausibility of the Local Origin of 10-TeV Bump \protect\label{subsec:PlausibilityOfLocal}}

\subsection{Rigidity Spectrum}

\begin{figure}[tb!]
	\includegraphics[viewport=100bp 0bp 842bp 590bp,clip,scale=0.38]{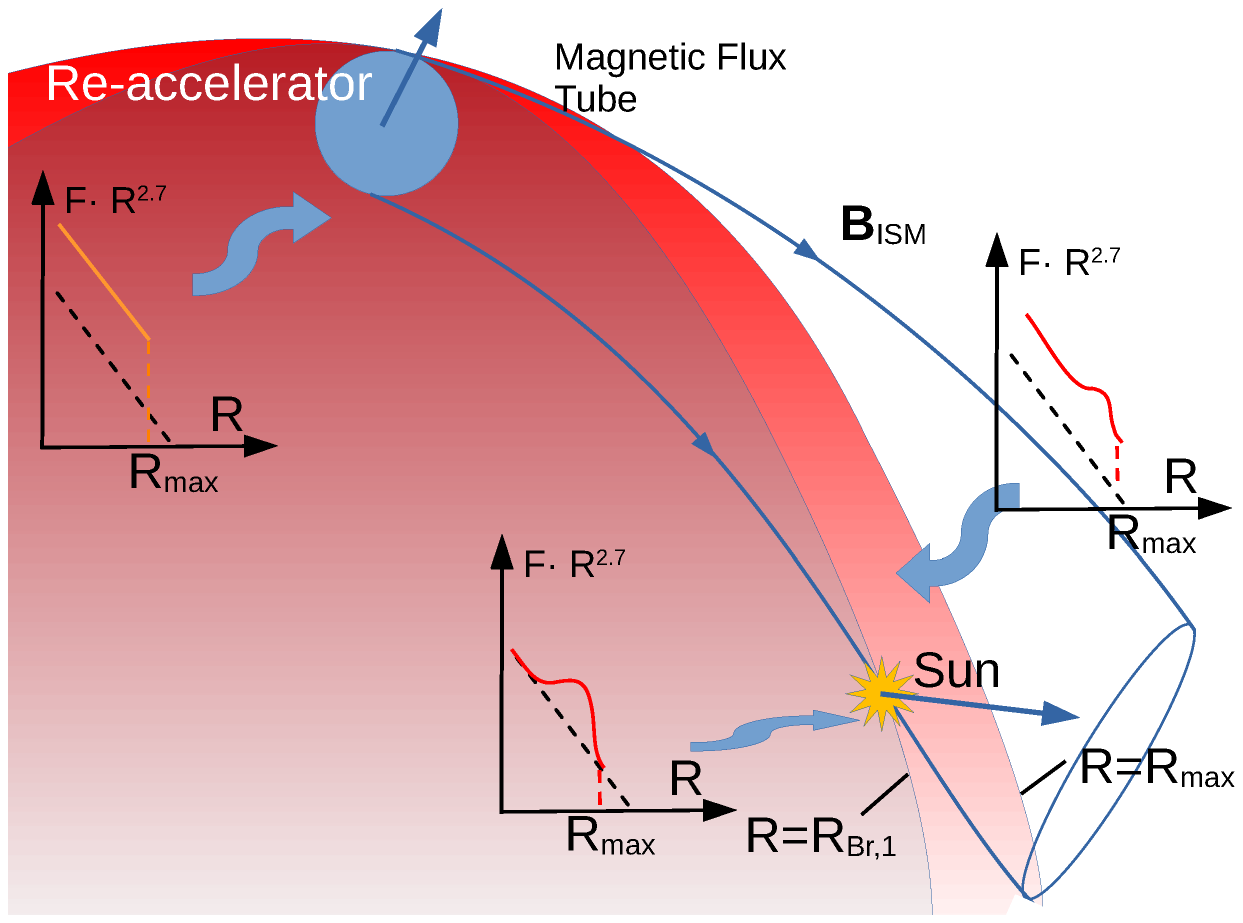}
	\caption{Schematics of a moving CR reaccelerator (e.g., a stellar bow shock--wind
		termination shock) and reaccelerated CRs that diffuse predominantly
		along the field lines, thus forming a wake behind this object. Three
		types of  reaccelerated CR spectra are shown at different positions
		in the wake. Dashed black lines show the CR background spectrum in
		each of the three insets schematically showing the observed total
		spectra (solid lines),  depending on the observer's position in the
		stellar wake (see text). The edge of the broader shaded area marks
		the boundary which the reaccelerated particles with $R=R_{\text{max}}$
		can reach, while those with $R<R_{\text{max}}$ can not. Outside of
		the inner shaded area, particles with $R<R_{\text{1}}$ cannot be
		observed. The Sun is located at this edge.}
	\label{fig:BowShocks} 
\end{figure}

Suppose a stellar bow shock is moving through the ISM, Fig.\ \ref{fig:BowShocks}.
At any instant, it populates a particular magnetic flux tube with
CRs reaccelerated up to $R_{\text{max}}$. This quantity is limited
by the requirement that Larmor radii of these particles $r_{g}=cR_{\text{max}}/B$
do not exceed the bow-shock size. The latter may extend up to $\sim10^{4}$
au \citep{Wood2002}. An enhanced magnetic field blown out of the
stellar cavity and compressed around the apex of the bow shock may
significantly ease this constraint. Particles, reaccelerated at the
bowshock and the stellar wind termination shock then propagate primarily
along the tube to a distance increasing with their rigidity. While
moving further through the ISM, the star leaves a turbulent wake of
flux tubes filled with reaccelerated CRs. It continues to broaden
along and across the field with the strogly different diffusivities $\kappa_{\parallel}\gg\kappa_{\perp}$. 
These quantities are related to the ISM frame, assuming
that there is no bulk fluid motion in the wake. At any given distance
behind the star, the breadth of the wake depends on the particle rigidity.
Therefore, depending on the position of the observers in the wake,
they will see the reaccelerated particles in the range $R_{\text{1}}<R<R_{\text{max}}$.
Here, $R_{\text{max}}$ is the maximum rigidity attainable by the
reacceleration. At the sun position, particles with $R<R_{\text{1}}$
(the first break, Fig.\ref{fig:Bump}) have not reached the observation
point yet, as they diffuse slower. As mentioned above, the wake also
has its thickness in the direction across the plane fixed by the star
velocity and magnetic field vectors. Here, we ignore it by assuming
that the sun is well inside the wake in this direction.

Assume further the Sun is in the wake where only particles with rigidities
$R\ge R_{1}$ can reach (the first break, see Fig.\ \ref{fig:BowShocks}).
We, therefore, see the spectrum flattening at $R>R_{1}$. The {\em
Sun's position in the wake and the energy upshift due to the reacceleration}
shape the bump appearance (shown in the bottom spectrum in the
figure). These model parameters fix its magnitude (factor $\approx$$2.4$
above the background) and its rigidity ($\sim$$10$ TV), while the
details of reacceleration are less critical. We can get a glimpse
into the spectrum variations at higher rigidities in the flux tube
using Eq.~(\ref{eq:SolFinal}) below. In particular, the second break
at $R=R_{2}$ (close to the maximum at $\approx$$10^{4}$GV in Fig.\ref{fig:Bump})
forms where the diffusion of reaccelerated CRs along the tube exceeds
their convection with the ISM flow out of the tube. The spectrum slope
becomes close to that it has at and behind the shock surface. At yet
higher rigidities it approaches that of the CR background, especially
when the CR losses out of the tube dominate.

The acceleration depends on the local angle between the shock normal
and the magnetic field: quasi-perpendicular shock ($\vartheta_{nB}\approx\pi/2$),
vs.\ quasiparallel shock ($\vartheta_{nB}\ll1$). A dull cone-shaped
stellar bow shock makes a wide range of angles to the local field,
$\vartheta_{nB}\le\pi/2$, with a possible exception for $\vartheta_{nB}\ll1$.
Even a weak quasi-perpendicular shock ($\vartheta_{nB}\approx\pi/2$)
of a size exceeding the particle Larmor orbit can provide the necessary
energy upshift, operating in a shock-drift acceleration (\textbf{SDA})
regime. All it takes is that the shock overruns a significant portion
of the particle's Larmor orbit before the particle escapes the shock
front along the field line. A single complete orbit crossing takes
$r_{g}/U_{\text{shock}}\approx30$ yrs for $r_{g}\sim3\cdot10^{15}$
cm (a 10 TV proton in a 10 $\mu$G field) and $U_{\text{shock}}\approx30$
km/sec. Note that direct Voyager 1, 2 measurements give 7--8 $\mu$G
just outside the heliosphere \citep{Burlaga2019}. As mentioned earlier,
stars with 
winds more powerful than solar, such as the $\epsilon$ Eridani star
discussed in \citetalias{MalkMosk_2021}, likely have a stronger field
outside the astrosphere due to the Axford-Cranfill effect. It constitutes
a magnetic pillow made of the stellar magnetic field blown out by
the wind.

The reacceleration requirements are relatively mild. The bump height
with respect to the underlying background spectrum at $R\approx10$ TV
is a factor of 2.4. However, because the background spectrum is relatively
steep ($\gamma_{b}\approx2.85$), it suffices to reaccelerate the
background particles with the initial rigidity $R\approx10/1.4\approx7.1$
TV to $R\approx10$ TV, which will elevate the level at $10$ TeV
by the required factor of $2.4$. A particle energy gain after its Larmor circle orbit is crossed by
a shock can be calculated from the condition of adiabatic compression:
$p_{\perp}^{2}/B=const$, where $p_{\perp}$ is the perpendicular
to the magnetic field component of particle momentum. Therefore, even
a factor of 2-3 shock (magnetic field $B$) compression would suffice
to make the bump visible. The momentum gain should further increase
due to the Axford-Cranfill effect. It should be noted that the wake
itself is formed primarily by the shocked stellar wind blown by a
star with a mass loss rate at least 30 times the solar one. The flow
past the termination shock is likely to be highly turbulent, thus
offering additional mechanisms of reacceleration. They are worth a
separate study. 

\begin{table*}[tb!]
\centering \caption{Model parameters and fit results for the proton spectrum.\protect\label{tab:Model-parameters-and}}
\begin{tabular}{lccc}
\hline 
Parameter (St.\ err.\ \%)  & $R_{0}$(GV)  & $R_{L}$(GV)  & $q_{s}$ \tabularnewline
\hline 
Realistic Model (RM)  & 5878 (3.5\%)  & $2.24\times10^{5}$ (28\%)  & 4.20 \tabularnewline
Loss-Free Model (LF)  & 4794 (3.2\%)  & $\infty$  & 4.73 \tabularnewline
\hline 
 &  &  & \tabularnewline
\end{tabular}
\end{table*}

\begin{table*}[tb!]
\centering \caption{Input parameters for CR species derived from their LIS \citep{2020ApJS..250...27B,2021ApJ...913....5B}
at $\approx$100 GV.\protect\label{tab:Input-parameters-and}}
\begin{tabular}{lcccccc}
\hline 
Parameters  & H  & He  & B  & C  & O  & Fe\tabularnewline
\hline 
$A_{b}$ (m$^{-2}$ s$^{-1}$ sr$^{-1}$ GV$^{-1}$)  & $2.32\times10^{4}$  & 3631  & 70.2  & 111  & 108  & 11.6\tabularnewline
\hline 
$\gamma_{b}$  & 2.85  & 2.77  & 3.09  & 2.75  & 2.73  & 2.66\tabularnewline
\hline 
 &  &  &  &  &  & \tabularnewline
\end{tabular}
\end{table*}

Consider now an oblique portion of the bow shock to demonstrate  a
straightforward possibility of CR reacceleration. We will add to the
standard DSA scheme \citep{BlandOst78} losses from the flux tube,
which are  essential at higher rigidities, $R>R_{L}$. The result
of reacceleration can then be cast as \citepalias{MalkMosk_2021,MalkMosk2022}:
{\small
\begin{equation}
f_{\text{CR}}\left(R\right)=A_{b}R^{-\gamma_{b}}\left\{ 1+\frac{\gamma_{b}+2}{q_{s}-\gamma_{b}}\exp\left[-\left(\frac{R_{0}}{R}\right)^{a}-\sqrt{\frac{R}{R_{L}}}\right]\right\} .\label{eq:SolFinal}
\end{equation}
}Here $\gamma_{b}$ (e.g., $\approx$2.85 for protons) and $A_{b}$
are the \emph{known} background CR index and normalization factor,
respectively, $q_{s}$ is the shock spectral index $q_{s}=\left(r+2\right)/\left(r-1\right)$,
where $r$ is the shock \emph{unknown} compression, $R_{0}\propto\left(L_{s}W/\sqrt{l_{s}}\right)^{1/a}$
is the characteristic bump rigidity that depends on the \emph{unknown}
distance to the shock, $L_{s}$, its size, $l_{s}$, and the level
of turbulence in the flux tube, $W$, and, especially, its index $a$ (see below). The \textbf{\emph{three}}\emph{
fundamental unknown bump-specific CR fitting parameters $q_{s},\,R_{0},$
and $\,R_{L}$ (Table~\ref{tab:Model-parameters-and}), can be obtained
from fitting Eq.~(\ref{eq:SolFinal}) to the proton spectrum (Fig.\ \ref{fig:Bump}),
the best measured among CR species.} Therefore, our model needs only
three parameters, which is two times less than the required minimal
number of simple geometric parameters defining the bump shape, which
is equal to \textbf{six} \citepalias{MalkMosk_2021}. \textit{Moreover,
as the spectrum is given in the element-invariant rigidity form, the
same fitting parameters derived from the proton spectrum can be used
to predict the spectra of other elements, such as He, C, O, etc.,
and the secondaries with }\textbf{\textit{no free parameters}}\textit{
\citepalias{MalkMosk2022}. Each CR species has its fixed known input parameters
$A_{b}$ and $\gamma_{b}$ derived from their local ISM spectra below
the bump, different from those of CR protons, see Sect.\ \ref{HeCO}.}

However, the most critical parameter for the fit is the turbulence-related
index $a$ in the flux tube: $W_{k}\propto k^{a-2}$, where $W_{k}$
is the wave spectral density. For calculations shown in Fig.\ \ref{fig:Bump},
we have determined it from dimensional reasoning to be $a=1/2$, which
corresponds to the Iroshnikov-Kraichnan (IK) turbulence model, $k^{-3/2}$.
Still, as we have used only three fitting parameters in Eq.~(\ref{eq:SolFinal})
out of the six that are required for a full bump characterization,
we can verify our calculation of $a$ by minimizing the data mismatch.
The mismatch has a very sharp minimum near $a=1/2$ \citepalias{MalkMosk_2021}.

\begin{figure}[b!]
	\includegraphics[scale=0.16]{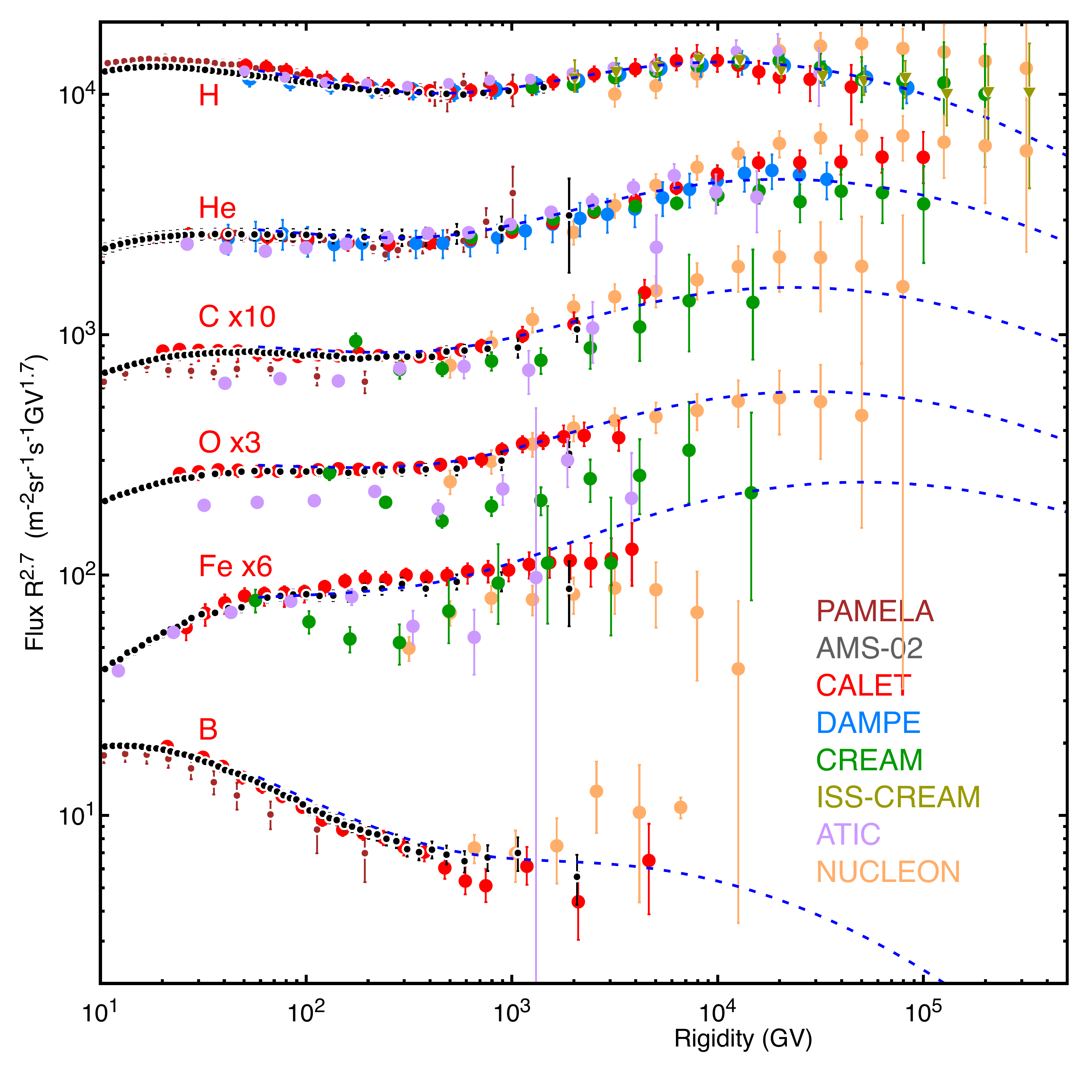} \caption{Spectra of CR species: protons, He, B, C, O, Fe. The compilation of
		data includes AMS-02 \citep{Aguilar2021,2021PhRvL.126d1104A}, ATIC
		\citep{2009BRASP..73..564P}, CALET \citep{CALET_2019PhRvL,2020PhRvL.125y1102A,2021PhRvL.126x1101A,AdrianiCalet2022,2022PhRvL.129y1103A,2023PhRvL.130q1002A},
		CREAM \citep{2009ApJ...707..593A,2017ApJ...839....5Y}, DAMPE \citep{Dampe2019a,2021PhRvL.126t1102A},
		ISS-CREAM \citep{ISS-Cream2022}, NUCLEON \citep{2020JETPL.111..363K,2020PhLB..81135851K,2019AdSpR..64.2546G},
		and PAMELA \citep{2011Sci...332...69A} instruments. 
		The proton spectrum is plotted as in Fig.~\ref{fig:Bump}. CALET
		B, C, O, Fe data are renormalized up by a factor of 1.3 to match the
		AMS-02 data, while the CALET He data are not adjusted. The blue dashed
		lines show the local \emph{ISM} spectra Eq.~(\ref{eq:SolFinal})
		in the realistic model with losses \citepalias{MalkMosk2022}, Tables~\ref{tab:Model-parameters-and}
		and \ref{tab:Input-parameters-and}.}
	\label{fig:CR_Spectra} 
\end{figure}

Moreover, the fit presented in Fig.\ \ref{fig:Bump} for $a=1/2$
\citepalias{MalkMosk_2021} rules out other popular turbulence models,
entertained in CR acceleration and propagation studies, such as $1/k$
(Bohm model) and $k^{-5/3}$ (Kolmogorov model). In particular, the
Kolmogorov and Goldreich-Shridhar spectra do not fit the new high-fidelity
data, shown in Fig.\ \ref{fig:Bump}. These spectra will likely dominate
the ISM turbulence outside the flux tube. The fit does not call them
into question since the IK spectrum is derived for a specific CR driver
in the flux tube. Even more significant deviation from the data gives
the Bohm turbulence regime, which, in turn, is expected closer to
the shock that reaccelerates CRs but does not contribute to their
propagation over the entire tube length. Our results thus point to
the CR-filled flux tube (or its wake) from a set of choices for explaining
the 10-TeV bump phenomenon. It most likely connects the Sun with a
CR-reaccelerating \emph{shock} 3-10 pc away.

\subsection{Heavier CR species}
\label{HeCO}

According to Eq.~(\ref{eq:SolFinal}), the CR species are differentiated
solely by their spectral index $\gamma_{b}$ below the first break
$R_{1}$, which \emph{is fixed by the data.} We derive $\gamma_{b}$
from a fit to the LIS by \citet{2020ApJS..250...27B,2021ApJ...913....5B}
at $R\approx$100 GV (Table\ \ref{tab:Input-parameters-and}). On
one hand, this is significantly below the break rigidity $R_{1}$
and is thus representative for the background spectrum. On the other
hand, the solar modulation at such rigidity is fairly weak. Meanwhile,
a straight fit to the data would produce very similar values.

We illustrate this in Fig.\ \ref{fig:CR_Spectra} that shows proton,
He, B, C, O, and Fe spectra. The proton spectrum is the same as in
Fig.~\ref{fig:Bump}. CALET B, C, O, Fe data are renormalized up
by a factor of 1.3 to match AMS-02 data, while CALET He data \citep{2023PhRvL.130q1002A}
and data by other instruments are not adjusted. Given almost the same
spectral indices for He, C, O below the first break $R_{1}$, their
spectral shapes are identical. We also show the spectra of B and Fe,
which have different spectral indices below the first break $R_{1}$.
Spectra of all species show good agreement with available data. Good
quality data for B and Fe around the second break $R_{2}$ are desired,
but not available yet due to the low flux and thus limited statistics.
When available in the future, they could be used to discriminate between
models.

\subsection{Anisotropy\protect\label{subsec:Anisotropy2}}

If the \emph{shock is located at} a few particle's mean free paths
from the observer, a sharp increase in the CR intensity across the
magnetic equator is expected and was indeed observed in \citet{Abeysekara2019},
Fig.\ \ref{fig:Anisotropy}.

\begin{figure}[tb!]
	\includegraphics[scale=0.42]{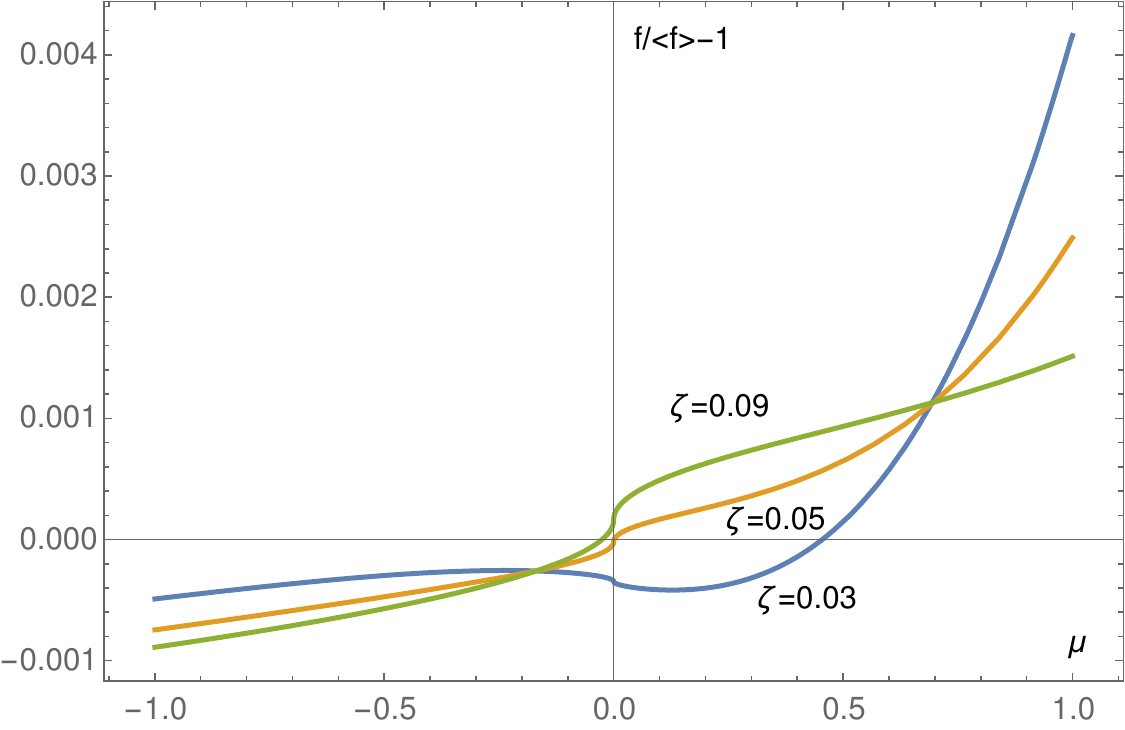} \caption{Three solutions as a function of $\mu$ shown as a relative anisotropy,
		$\left(f-\left\langle f\right\rangle \right)/\left\langle f\right\rangle $.
		The isotropic part $\left\langle f\right\rangle \approx0.5$0 for
		all three values of $\zeta$, and decreases with $\zeta$ insignificantly,
		since $\nu_{\perp}=10^{-6}$ and $\lambda_{0}\approx$0.00447, $\lambda_{1}\approx45.0$,
		$\lambda_{2}\approx132$. The source anisotropy  $Q_{1}\left(\chi\right)=1+0.01\left|\chi\right|^{3}$,
		introduced by writing the source $Q$ in eq.(\ref{eq:FPinit}) as
		$Q=\nu_{\text{IK}}Q_{0}\left(p\right)Q_{1}\left(\mu\right)$.  Eigenmodes
		decay with distance as $f_{n}\propto\exp\left(-\lambda_{n}\left|\zeta\right|\right)$.}
	\label{fig:AnisInChiMu} 
\end{figure}

We use the Fokker-Planck equation for the CR propagation along the
field \citepalias{MalkMosk_2021}: {\small
\begin{equation}
c\mu\frac{\partial f}{\partial\zeta}+\nu_{\perp}^{\prime}f-\frac{\partial}{\partial\mu}\left(1-\mu^{2}\right){\mathcal{D}}\left(p,\mu\right)\frac{\partial f}{\partial\mu}=Q\left(p,\mu\right)\delta\left(\zeta\right).\label{eq:FPinit}
\end{equation}}
\noindent Here $\mu$ is the pitch angle cosine of a particle with momentum
$p$. The spatial coordinate $\zeta$ is directed along the field,
and the source is at the origin. The other two coordinates are removed
by averaging the distribution function across the flux tube. The particle
flux through its boundary enters as $\nu_{\perp}^{\prime}f$ on the
l.h.s. The CR scattering frequency $\mathcal{D}$ is related to thered
Alfv\`en wave fluctuations $E_{k}^{A}$, by an IK spectrum. The solution
is found by decomposing it in a series of eigenfunctions of the operator
on its l.h.s. Starting from some distance from the source, it is dominated
by the lowest eigenvalue $\lambda_{0}\approx2\sqrt{5\nu_{\perp}}$
and can be approximated near magnetic equator as $f_{0}\approx\sqrt{5/\lambda_{0}}\left[1+\left(\lambda_{0}/4\right)\left|\mu\right|^{1/2}\text{sgn}\mu\right]$,
thus showing a sharp increase over the magnetic equator ($\mu=0$),
consistent with the data presented in Fig.\ \ref{fig:Anisotropy}. Here, we have introduced the loss rate $\nu_{\perp}=\left|\mu\right|^{1/2}\nu_{\perp}^{\prime}$/$\nu_{\text{IK}}$,
assuming it $\mu$- independent. The Iroshnikov-Kraichnan scattering
rate is, in turn, related to the pitch-angle scattering rate $\mathcal{D}$
in eq.(\ref{eq:FPinit}) as follows: $\mathcal{D}=4\nu_{\text{IK}}\left|\mu\right|^{1/2}$.
Further details of the analysis of this equation can be found in the
above reference. Note that the Kolmogorov turbulence would produce
a steeper rise of CR intensity across the magnetic equator, $\propto|\mu|^{1/3}\text{sgn}\mu$,
which might become discernible from the IK turbulence if the HAWC-IceCube
statistics improve. At the same time, an isotropic pitch angle scattering
that results in a Bohm diffusion regime would produce no such effect
at all, as the scattering frequency is analytic ($\mathcal{D}\left(\mu\right)\approx const$)
at $\mu=0$. Besides, this can be seen from an exact solution of the
underlying Fokker-Planck equation \citep{malkov2017exact}. We also
note here that the three turbulence models discussed above are abundantly
present in the solar wind \citep{Chen_2020}.

The small-scale anisotropy ($f_{n\ge1}$) can also be observed at
the dimensionless distance to the source $\left|\zeta\right|\la1/\lambda_{1}$,
along with the sharp increase across the magnetic equator discussed
in the preceding paragraph. To illustrate the variation of small-
and large-scale anisotropy with the distance, we show in Fig.\ \ref{fig:AnisInChiMu}
the angular distributions of CRs at three different dimensionless
distances $\zeta$ (given in units of $c/\nu_{\text{IK}}$). At a
short distance ($\zeta=0.03$), there is a strong increase in the
field-aligned particles ($\mu\approx1$), but the curve is inconsistent
with the observed map (Fig.\ref{fig:Anisotropy}) at $\mu=0$. For
large distances ($\zeta=0.09$), the behavior at $\mu=0$ is correct,
but there is not significant enhancement at $\mu=1$, observed in
$B_{\text{LIMF}}$ direction. Only at $\zeta=0.05$, both effects
are consistent with the observations. We thus see that when $\zeta$
decreases by a factor of 3 (from 0.09 to 0.03), the CRs dipolar distribution
shrinks to a sharp one toward the source. It shows a progressively
stronger CR field alignment. Along with a jump across the magnetic
equator at $\mu=0$, contained in $f_{0}$, an enhancement at $\mu\approx1$,
associated with $f_{1}$, is observed at the distances $0.05<\zeta<0.1$.
This angular pattern can be found on the map of CR arrival directions
in Fig.\ \ref{fig:Anisotropy}. Its interpretation based on the $\epsilon$-Eridani
proximity naturally explains the puzzle of CRs predominantly arriving
from the Galactic ANTI-center direction. At the same time, \emph{$\gamma$-ray
observations of the diffuse emission by Fermi}-LAT testify to a higher
concentration of CRs in the inner Galaxy \citep{Ackermann2012,zhang2023galactic},
see also \citet{Gabici:2023/V}. By contrast, even an advanced propagation
model \citep{ChernyshovIvlevDogiel2023} based on distributed CR sources
is unlikely to reproduce the sharp, especially Galactic anti-center
prevalent CR anisotropy. We will discuss the alternative models in
Sec.\ref{sec:Alternative-Explanations-of}.

\subsection{$\gamma$-rays from $\epsilon$ Eridani}

Interestingly, there is an indication of the $\gamma$-ray emission
in the direction of the proper motion of $\epsilon$ Eridani (RA:
$-974.758$ mas/yr, Dec: 20.876 mas/yr, \citealt{GaiaCollaboration2020})
at a distance of $\la$0.5$^{\circ}$, which corresponds to the size
of the astrosphere and the location of the bow shock. Fig.\ 3 in
\citet{Riley2019} shows the position of $\epsilon$ Eridani with
the white circle in the middle of the $10^{\circ}\times10^{\circ}$
region of interest. An excess emission to the right of the star (the
direction of the proper motion) is marked with a green circle \emph{partially}
overlapping with the white circle, but the spectrum of the excess
emission is not provided. The soft spectrum of the emission from \emph{the
star} may suggest background contamination rather than reflecting
emission from the star itself. \citet{Riley2019} speculate about
the emission from interactions of cosmic rays with the debris disk
around the star or the stellar activity. Unfortunately, the angular
resolution of \emph{Fermi}-LAT in the analyzed energy range 100 MeV
-- 3.16 GeV is \emph{rather} poor and does not allow us to reconstruct further details. 

\section{Alternative Explanations of the 10-TeV CR Bump\protect\label{sec:Alternative-Explanations-of}}

The now firmly established 10-TeV bump phenomenon has been emerging
for more than a decade following improvements in observations. The
first reports on spectral hardening in the p-He spectra at a few 100
GV date back to the mid-2000s (for an abbreviated list of references see, e.g., \citetalias{MalkMosk_2021}). The full appreciation of its
realism and significance was gained only a few years ago, after the
accuracy of the spectral data was significantly improved. This allowed
the AMS-02 \citep{2015PhRvL.114q1103A} and CALET \citep{CALET_2019PhRvL}
teams to quantify the spectral hardening by measuring the sharpness
of the transition in the proton spectrum, $s=0.024\pm_{0.021}^{0.034}$
and $0.089\pm0.133$, correspondingly, using the fitting function:
\begin{equation}
F_{\text{CR}}\propto R^{-\gamma}\left[1+\left(R/R_{\text{br}}\right)^{\Delta\gamma/s}\right]^{s}.\label{def:s}
\end{equation}
Here $\gamma$ is the spectral index below the break, and $\Delta\gamma$
is the value of the spectral break. The sharpness of the break was
also confirmed with larger statistics, $s=0.09\pm_{0.03}^{0.04}$
\citep[AMS-02,][]{Aguilar2021}. Further critical insight has been
gained from more precise measurements of other elements, especially
the secondary ones, as well as from the detailed angular distribution
of the protons around 10 TeV.

At first, the emerging discovery was perceived as an isolated spectral
hardening and triggered suggestions of two unrelated mechanisms for
its explanation. The first mechanism was a straightforward combination
of two or more independent CR sources, such as nearby SNRs, including
the CR background power-law spectrum. With significant error bars
attached to the data sets, these models were consistent with
the shape of the primary CR spectra. However, persuasive arguments
\citep{VladimirMoskPamela11} have promptly pointed to a likely inconsistency
of these models with even more substantial hardening observed in the
spectra of secondary CR species. If the primary nuclei produced them
after being accelerated in the nearby sources, an unrealistically
large amount of target material for the spallation reactions between
the source and the observer would be necessary.

The second mechanism seemingly circumvents the problem with secondaries
by invoking a presumed break in the turbulence spectrum that controls
particle transport \citep{VladimirMoskPamela11}. The recent review
of this mechanism and the list of references can be found in \citet{Blasi2022},
whereas interesting new approaches were published afterward \citep{Chernyshov2022,ChernyshovIvlevDogiel2023}.
According to the models built on this premise, the break in the turbulence
spectrum at the wave number $k_{\text{br}}$ directly translates into
the break in the momentum spectrum of propagating particles at $p_{\text{br}}=eB/ck_{\text{br}}$,
because of this cyclotron resonance condition. Breaks in the ISM turbulence
spectra are possible, and such breaks are even observed in the solar
wind \emph{in situ} \citep{Chen_2020}. However, there are three aspects
of particle propagation not included in these models, whereby the
turbulence break \emph{will not be mirrored} in the observed \emph{sharp
break} in the particle momentum spectrum. We describe them below.

\subsection{The Spectral Break is too Sharp to form Remotely\protect\label{subsec:Spectral-Break-Broadening}}

Were the observed CRs propagated a significant distance to us,  a
sharp spectral break  would be smeared out over the momentum range
$\Delta p\gtrsim p_{\text{br}}$, as we have argued at the beginning
of Sec.\ref{sec:Solar-Neighborhood-Origin}.  The observed break requires
the opposite condition $\Delta p\ll p_{\text{br}}$ that indicates
a closeby source responsible for the break. In fact, the parameter
$\Delta p/p_{\text{br}}$, whose numerical equivalent was denoted
by $s$, Eq.\ (\ref{def:s}), was directly obtained from the observations
(also shown in Fig.\ \ref{fig:Bump} above) being $s\approx0.1$.
It should be noted here that  the expression in eq.(\ref{def:s})
is \emph{ad hoc}, unlike Eq.~(\ref{eq:SolFinal}). The latter results
from solving a particle reacceleration problem and is used for the
fit shown in Fig.\ \ref{fig:Bump}. Nevertheless, the above formula
for $F_{\text{CR}}$ in Eq.\ (\ref{def:s}) provides a helpful tool
for ruling out a broad range of models, including those discussed
at the beginning of this section.

We illustrate this point in Fig.~\ref{fig:break}. It shows fits
to the AMS-02 CR proton data \citep{Aguilar2021}. The original fit
\citep[45 GV--1.8 TV,][]{Aguilar2021} with functional dependence
given in Eq.~(\ref{def:s}) is shown by the  black solid line, where
the numerical values of parameters are slightly adjusted. Such a fit
requires a fairly small value of $s\approx0.09$. The latter means
that the two power laws, at low and high energies (solid green lines),
should have a sharp cutoff near the break rigidity. Otherwise, a simple
sum of the two power-laws, the combined spectrum ($s=1$), is too
broad to fit the data. The sum of these power-laws  is shown by the
blue solid and dotted lines for different normalizations. 

\subsection{Artificial Resonance Sharpening Produces the Break Artifact\protect\label{subsec:Artificial-Resonance-Sharpening}}

A sharp break in the turbulence spectrum produces only a smooth variation
of the power-law index in the particle momentum spectrum during their propagation through the turbulent ISM. 
This relation between the two breaks, which follows from
the character of particle interactions with resonant magnetic perturbations, is often overlooked in the literature.
 
Consider a circularly polarized Alfven
wave with the wave number $k$ propagating along the magnetic field. This setting is most favorable and straightforward for connecting the breaks in wave and particle spectra directly.
In the wave's rest frame, a particle spiraling around the unperturbed field
with an average velocity $v_{\parallel}$ perceives an oscillatory
force from the wave of the frequency $kv_{\parallel}\equiv kv\mu$,
where $\mu$ is the cosine of the particle's pitch angle and $v\approx c$. An irreversible
pitch angle variation occurs when the particle passes a cyclotron resonance that, for
ultrarelativistic particles, can be written as $p=eB/ck\mu$. It involves
one wave parameter, $k$, and two parameters for the particles, $p$
and $\mu.$ 

However, typical models that associate the wave spectral breaks with
those of the particles use a simplified resonance condition, $p\approx eB/ck$,
involving one-to-one wave-particle resonant association. This simplification
was introduced by \citet{Skill75c} out of ``algebraic convenience''
and called ``resonance sharpening.'' While broadly used and helpful
in studying smooth particle and wave spectra, it is inadequate for
treatments of broken ones, especially the sharply broken spectra discussed
above. Indeed, it effectively collapses a nearly isotropic particle distribution in $\mu$ to a delta-like distribution at some $\mu=\mu_0\simeq 1$. To demonstrate the inadequacy of this approximation for sharply broken wave spectra,
consider two particles with different momenta $p_{1}$
and $p_{2}>p_{1}$. According to the ``sharpened'' resonance conditions,
each of them interacts with a unique wave with $k=k_{1}$ and $k=k_{2}<k_{1}$.
Therefore, if there is a break in the wave spectrum somewhere between
$k_{2}$ and $k_{1}$, a corresponding break is expected to form in the
particle spectrum between $p_{2}$ and $p_{1}$.

If the exact resonance condition, $p=eB/ck\mu$, is applied instead,
two particles with momentum $p_{1}$ but with different $\mu$ may
interact with both $k_{1}$ and $k_{2}$ waves, which is also true
for the particles with momentum $p_{2}$. Therefore a break in the
wave spectrum will not profoundly affect the particle spectrum. To
put it formally, consider the wave-particle interaction term in a quasilinear approximation, which in a particle kinetic
equation appears in the form $\int dk\delta\left(p\mu-eB/ck\right)W_{k}(\dots)$.
Here $W_{k}$ is the wave energy density.  We  denoted components
of the matrix element of the wave-particle interaction unessential
for this discussion by $(\dots)$.  Eventually, pitch-angle scattering translates into particle diffusion. The argument
behind the formation of the spectral break in the particle spectrum
is that the break should form in the particle diffusion coefficient
in the first place. However, this does not happen, which can be readily
seen from the following quasilinear result for the particle
diffusion coefficient \citep{KulsrNeutr69}:

\begin{figure}[tb!]
	\includegraphics[scale=0.172]{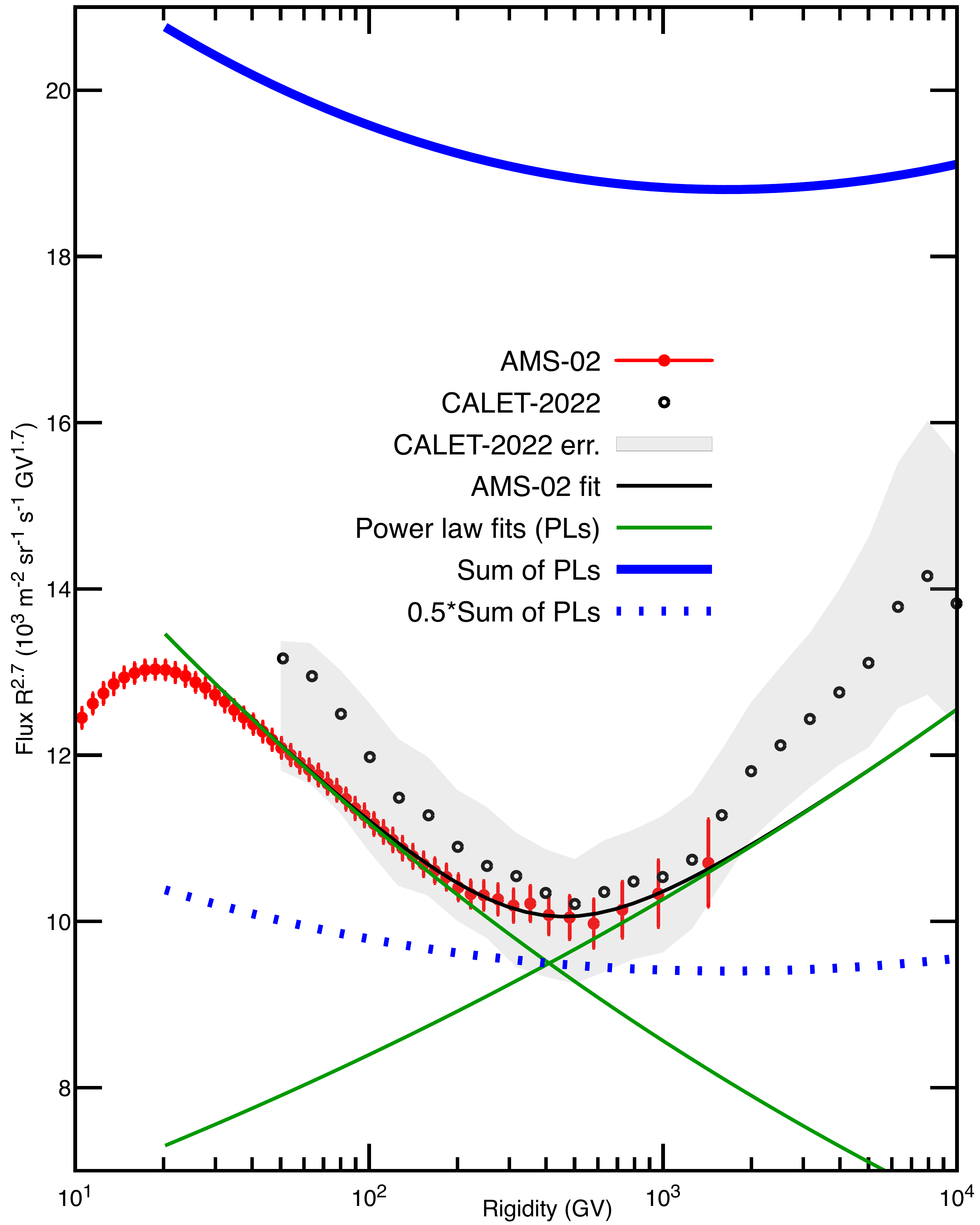} \caption{Illustration of fits to the AMS-02 CR proton data \citep{Aguilar2021}.
		The original fit \citep{Aguilar2021} with functional dependence given
		in Eq.~(\ref{def:s}) is shown with the black solid line, where the
		numerical values of parameters are slightly adjusted. For clarity, only CALET data at high energies are shown. 
		The asymptotic
		power laws at low ($\propto R^{-\gamma}$, $\gamma\approx-2.8$) and
		high ($\propto R^{-\gamma+\Delta\gamma}$, $\Delta\gamma\approx0.2$)
		rigidities derived from Eq.~(\ref{def:s}) are shown by the green
		lines. The solid blue line represents their sum, while the dotted
		one shows a half of that, for comparison. It is clear that no curve
		in between would match the composite AMS-02/CALET data. }
	\label{fig:break} 
\end{figure}

\begin{equation}
\kappa_{\parallel}\left(p\right)=\frac{cB^{2}p^{2}}{m^{2}\omega_{c}^{2}}\int\frac{\left|\mu\right|\left(1-\mu^{2}\right)d\mu}{W_{k=eB/cp\mu}}.\label{eq:pardif}
\end{equation}
Representing the integral as

{\small
\[
\int\frac{p\mu^{2}\left(1-\mu^{2}\right)d\mu dk}{W_{k}}\delta\left(kp\mu-\frac{eB}{c}\right)=\int\frac{\mu_{k}^{2}\left(1-\mu_{k}^{2}\right)dk}{kW_{k}},
\]}
\noindent where $\mu_{k}\equiv eB/cpk$, reveals any kink (discontinuity in
a derivative) in the turbulence spectrum $W_{k}$ being smoothed out
after the integration over $k$. The discontinuity of $\partial W_{k}/\partial k$
at $k=k_{\text{br}}$ will then translate only in the discontinuity
of $\partial^{2}\kappa_{\parallel}/\partial p^{2}$ at $p=p_{\text{br}}$,
thus rendering no break in $\kappa_{\parallel}\left(p\right)$ and
a correspondingly smooth particle spectrum. Asymptotically, at $p\gg p_{\text{br}}$,
the particle spectral slope will accommodate to the different slope
of $W_{k}$ at $k<k_{\text{br}}.$ However, the transition will occur
over a broad region $\Delta p/p_{\text{br}}>1$, contrary to the observations.
On a positive note, this gradual spectral flattening is consistent
with the CR observations in the knee area at 
PeV energies, thus having some potential for describing the behavior
of the CR spectrum over the momentum scales 1 TeV -- 1 PeV, but clearly
inconsistent with sharp spectral variations in the TeV range.

\subsection{Effect of Nonlinear Landau Damping on Sharp Spectra\protect\label{subsec:Effect-of-Nonlinear}}

Another common element of the models that attempt to translate a break
in the turbulence spectrum to that in the particle momentum spectrum
is the so-called nonlinear Landau damping. From the wave perspective,
this is the process of wave scattering on the plasma thermal particles:
$\omega,\boldsymbol{k}\to\omega^{\prime},\boldsymbol{k}^{\prime}$
and hence, another name for the process is ``induced scattering''
\citep{SagdGal69,Lee1973}. From the particle perspective, it describes
the thermal particle scattering on a beat wave with the frequency
$\omega\pm\omega^{\prime}$ and wave vector $\boldsymbol{k}\pm\boldsymbol{k}^{\prime}$.
This process is essentially a higher-order approximation in the expansion
of the wave-particle interaction in the wave amplitude, following
the standard quasilinear treatment and resulting in $\sim W_{k}^{2}$
terms in the wave kinetic equation. Given the linear dispersion of
long Alfv\`en waves, $\omega=k_{z}V_{A}$, it is clear that Alfv\`en waves
from a broad $k$-range generate beat waves with phase velocities
resonant with the thermal particles. It follows then that the concept
of ``resonance sharpening'' is even less suitable for the induced
scattering than for the lower order approximation described above.
In most of the models addressing spectral hardening, this process
is included in an oversimplified, integrated form that does not adequately
describe the particle resonance with beat waves (cf. \citealt{Chernyshov2022}
and \citealt{Lee1973}).

A more general smoothing effect on any possible break in the CR momentum
spectrum created in a significant volume of galactic disc and adjacent
halo is the following. Assume that despite the above arguments, the
CR break is created at some $p=p_{\text{br}}$. As it must be produced
by a change in the particle transport regime (diffusion, convection,
field-line streaming, and escape), the value of $p_{\text{br}}$ must
depend on the local parameters which vary considerably across the
galaxy volume. So the momentum derivative of the broken power law
will be $\partial F_{\text{CR}}/\partial p\propto F^{\prime}\left(p,\boldsymbol{r}\right)+\alpha\left(p,\boldsymbol{r}\right)\Theta\left[p-p_{\text{br}}\left(\boldsymbol{r}\right)\right]$,
where $\Theta$ is the Heaviside unit function and $F^{\prime}$ and
$\alpha$ are continious functions. Since we are observing
particles that come to us from at least a kpc distance, we must average this result over the respective volume. 
The averaging will smear out the jump in the Heaviside function and, therefore,
the break in $F_{\text{CR}}$.

To conclude this section, we reiterate that, in addition to the above arguments,  the second-order Fermi reacceleration
must smear out {\emph {any sharp break}} formed at distances longer than
a few hundred parsecs. The neutral gas damping of the ISM turbulence is
unlikely to suppress the reacceleration since the Local Buble plasma
($\sim$100 pc) is highly ionized. Only the local cloud network may
produce some damping effect, but it constitutes only an insignificant
(\textasciitilde{} 10 pc) fraction of the Local Bubble on the CR propagation path. 

\section{Conclusions and Discussion}

We have analyzed the high-fidelity data on the 10-TeV puzzling bump
in the Galactic rigidity spectrum. It likely originates from 3-10
pc of the Sun. A system of bow-plus-termination shocks associated
with a star, such as $\epsilon$ Eridani, possibly supplemented with
its notably powerful stellar wind, is capable of significant CR reacceleration.
Being magnetically connected with the Sun, the $\epsilon$ Eridani
system likely accounts for all observed bump features well within
the data uncertainties with only three free parameters derived from
a fit to the proton spectrum. A significant number of such features
discussed make a coincidental agreement \emph{rather} unlikely. Any
combination of primary CR sources, such as SNRs, and variations in
the CR propagation cannot account for the fine angular and energy
spectral structures. The resultant CR spectrum would be smoother
than observed since these sources are at least a few 100 pc away.

Observations of CRs in an intermediate energy range from 0.1--100
TeV are becoming increasingly instrumental in testing models for CR
acceleration and propagation, as this is their preferred territory,
void of difficulties associated with particles entering the acceleration
(injection of suprathermal particles) and escape after achieving the
highest energy possible for the presumed accelerators, such as SNR
shocks. Even more critical is the multifaceted character of the newest
observations, cutting across the rigidity, angular, and multispecies
aspects of the CR spectra in this range, thus helping us zero in on
the most tenable models. Regarding the model discussed in this paper,
we adhered to the following requirements: 
\begin{enumerate}
\item Fitting the bump rigidity spectrum, shown in the log-linear\footnote{For example, the AMS-02 data are not recommended to be shown in log-log
format, they deserve the log-linear scale (Prof.\ Samuel Ting, personal
communication, 2014).} format with the $R^{2.7}$ factor within minimal for a given rigidity
error bars, complementarily selected from the high-fidelity instruments. A formal
description of the bump requires six parameters (two rigidity values,
two index variations, and two sharpness parameters for each break
in the spectrum). The number of model parameters should thus be less
than 6. Our model uses just 3 parameters that encapsulate the distance/size
of the accelerator (shock), its strength, and particle losses from
the flux tube during their propagation to the observer.
\item Explaining both the sharp anisotropy in the bump CR arrival directions
across the local magnetic equator and its dipolar component. They
both point to the source located in the hemisphere opposite to the
Galactic center.
\item Explaining why the secondary CRs develop a somewhat stronger first
break than the primary CRs do. Note that the second break has not
been seen in the secondaries yet due to their soft spectra and thus
insufficient statistics at multi-TV rigidities, but this may change
soon.
\item Demonstration of intervening turbulence not smoothing out the bump
during the CR propagation from its source to the observer. Demonstration
of the turbulence index consistency with the bump shape.
\item Demonstration of magnetic connectivity with the source of CRs constituting
the bump; alternatively, the bump has to be devised to be generated
in a large volume of the Milky Way containing the heliosphere. This scenario would encounter insurmountable problems with the model requirements (1-4).
\end{enumerate}

\subsection{Challenge to the Model}

A concern may be raised about the $\epsilon$ Eridani bow shock--wind
termination shock system, which may be perceived as being too small
($\sim$$10^{4}$ au $\approx1.5\times10^{17}$ cm) for modifying
the CR spectrum around the 10-TeV energy to the observed bump amplitude
(proton gyroradius $\approx$$3\times10^{15}$ cm in 10 $\mu$G field).
In addition to the arguments made in Sec.\ \ref{subsec:PlausibilityOfLocal}
about a relatively weak acceleration required, this problem can further
be ameliorated by recent findings about the current Sun position relative
to the local clouds and the direction to $\epsilon$ Eridani. Namely,
once the CRs are reaccelerated at the $\epsilon$ Eridani, they must
pass to the Sun between the G-Cloud (which is in the Galactic center
direction as seen from the solar system) and the Local Interstellar
Cloud (LIC). Moreover, the Sun no longer resides in the LIC, as previously
believed, but is already in the LIC-G-cloud interaction medium \citep{Swaczyna2022}.
If so, the field extraneous to both clouds is then compressed by their
motion towards each other with a relative velocity $\approx$7.2 km/s.
The field was known to be tangent to both cloud surfaces \citep{Lallement2005},
thus most likely being, indeed, compressed by them. The compressed
intercloud field must create a magnetic focusing effect on the CR
passing the Sun, currently crossing the G-LIC interface. The magnetic
focusing should increase the CR flux through the flux
tube, thus facilitating the fit. Quantifying this increase merits a separate
study.

\section*{Acknowledgments}

We thank two anonymous referees for their helpful comments. Work of
MM, PD, and MC is supported by NSF grant AST-2109103, and, in part,
by NASA ATP 80NSSC24K0774. IM acknowledges partial support from NASA
grants 80NSSC23K0169 and 80NSSC22K0718.

\bibliographystyle{jasr-model5-names}
\bibliography{AdSR}

\begin{thebibliography}{52}
\expandafter\ifx\csname natexlab\endcsname\relax\def\natexlab#1{#1}\fi
\ifx\xfnm\relax \def\xfnm[#1]{\unskip,\space#1}\fi

\bibitem[{{Abdo} et~al.(2008){Abdo}, {Allen}, {Aune}, {Berley}, {Blaufuss},
  {Casanova}, {Chen}, {Dingus}, {Ellsworth}, {Fleysher}, {Fleysher},
  {Gonzalez}, {Goodman}, {Hoffman}, {H{\"u}ntemeyer}, {Kolterman}, {Lansdell},
  {Linnemann}, {McEnery}, {Mincer}, {Nemethy}, {Noyes}, {Pretz}, {Ryan},
  {Parkinson}, {Shoup}, {Sinnis}, {Smith}, {Sullivan}, {Vasileiou}, {Walker},
  {Williams} \& {Yodh}}]{Milagro08PRL}
\bibinfo{author}{{Abdo}, A.~A.}, \bibinfo{author}{{Allen}, B.},
  \bibinfo{author}{{Aune}, T.} et~al. (\bibinfo{year}{2008}).
\newblock \bibinfo{title}{Discovery of localized regions of excess 10-{TeV}
  cosmic rays}.
\newblock {\it \bibinfo{journal}{\prl}\/},  {\it
  \bibinfo{volume}{101}\/}\bibinfo{issue}{(22)}, \bibinfo{pages}{221101--+}.
  \DOIprefix\doi{10.1103/PhysRevLett.101.221101}.
  \href{http://arxiv.org/abs/0801.3827}{\tt arXiv:0801.3827}.

\bibitem[{Abeysekara et~al.(2019)Abeysekara, Alfaro, Alvarez, Arceo,
  Arteaga-Vel{\'a}zquez, Rojas, Belmont-Moreno, BenZvi, Brisbois, Capistr{\'a}n
  et~al.}]{Abeysekara2019}
\bibinfo{author}{Abeysekara, A.}, \bibinfo{author}{Alfaro, R.},
  \bibinfo{author}{Alvarez, C.} et~al. (\bibinfo{year}{2019}).
\newblock \bibinfo{title}{All-sky measurement of the anisotropy of cosmic rays
  at 10 {TeV} and mapping of the local interstellar magnetic field}.
\newblock {\it \bibinfo{journal}{\apj}\/},  {\it
  \bibinfo{volume}{871}\/}\bibinfo{issue}{(1)}, \bibinfo{pages}{96}.

\bibitem[{{Ackermann} et~al.(2012){Ackermann}, {Ajello}, {Atwood}, {Baldini},
  {Ballet}, {Barbiellini}, {Bastieri}, {Bechtol}, {Bellazzini}, {Berenji},
  {Blandford}, {Bloom}, {Bonamente}, {Borgland}, {Brandt}, {Bregeon},
  {Brigida}, {Bruel}, {Buehler}, {Buson}, {Caliandro}, {Cameron}, {Caraveo},
  {Cavazzuti}, {Cecchi}, {Charles}, {Chekhtman}, {Chiang}, {Ciprini}, {Claus},
  {Cohen-Tanugi}, {Conrad}, {Cutini}, {de Angelis}, {de Palma}, {Dermer},
  {Digel}, {Silva}, {Drell}, {Drlica-Wagner}, {Falletti}, {Favuzzi}, {Fegan},
  {Ferrara}, {Focke}, {Fortin}, {Fukazawa}, {Funk}, {Fusco}, {Gaggero},
  {Gargano}, {Germani}, {Giglietto}, {Giordano}, {Giroletti}, {Glanzman},
  {Godfrey}, {Grove}, {Guiriec}, {Gustafsson}, {Hadasch}, {Hanabata},
  {Harding}, {Hayashida}, {Hays}, {Horan}, {Hou}, {Hughes}, {J{\'o}hannesson},
  {Johnson}, {Johnson}, {Kamae}, {Katagiri}, {Kataoka}, {Kn{\"o}dlseder},
  {Kuss}, {Lande}, {Latronico}, {Lee}, {Lemoine-Goumard}, {Longo}, {Loparco},
  {Lott}, {Lovellette}, {Lubrano}, {Mazziotta}, {McEnery}, {Michelson},
  {Mitthumsiri}, {Mizuno}, {Monte}, {Monzani}, {Morselli}, {Moskalenko},
  {Murgia}, {Naumann-Godo}, {Norris}, {Nuss}, {Ohsugi} et~al.}]{Ackermann2012}
\bibinfo{author}{{Ackermann}, M.}, \bibinfo{author}{{Ajello}, M.},
  \bibinfo{author}{{Atwood}, W.~B.} et~al. (\bibinfo{year}{2012}).
\newblock \bibinfo{title}{{Fermi-LAT} observations of the diffuse
  {\ensuremath{\gamma}}-ray emission: Implications for cosmic rays and the
  interstellar medium}.
\newblock {\it \bibinfo{journal}{\apj}\/},  {\it
  \bibinfo{volume}{750}\/}\bibinfo{issue}{(1)}, \bibinfo{pages}{3}.
  \DOIprefix\doi{10.1088/0004-637X/750/1/3}.
  \href{http://arxiv.org/abs/1202.4039}{\tt arXiv:1202.4039}.

\bibitem[{{Adriani} et~al.(2019){Adriani}, {Akaike}, {Asano}, {Asaoka},
  {Bagliesi}, {Berti}, {Bigongiari}, {Binns}, {Bonechi} \&
  {Bongi}}]{CALET_2019PhRvL}
\bibinfo{author}{{Adriani}, O.}, \bibinfo{author}{{Akaike}, Y.},
  \bibinfo{author}{{Asano}, K.} et~al. (\bibinfo{year}{2019}).
\newblock \bibinfo{title}{Direct measurement of the cosmic-ray proton spectrum
  from 50 {GeV} to 10 {TeV} with the {Calorimetric Electron Telescope on the
  International Space Station}}.
\newblock {\it \bibinfo{journal}{\prl}\/},  {\it
  \bibinfo{volume}{122}\/}\bibinfo{issue}{(18)}, \bibinfo{pages}{181102}.
  \DOIprefix\doi{10.1103/PhysRevLett.122.181102}.
  \href{http://arxiv.org/abs/1905.04229}{\tt arXiv:1905.04229}.

\bibitem[{{Adriani} et~al.(2020){Adriani}, {Akaike}, {Asano}, {Asaoka},
  {Bagliesi}, {Berti}, {Bigongiari}, {Binns}, {Bongi}, {Brogi}, {Bruno},
  {Buckley}, {Cannady}, {Castellini}, {Checchia}, {Cherry}, {Collazuol},
  {Ebisawa}, {Fuke}, {Gonzi}, {Guzik}, {Hams}, {Hibino}, {Ichimura}, {Ioka},
  {Ishizaki}, {Israel}, {Kasahara}, {Kataoka}, {Kataoka}, {Katayose}, {Kato},
  {Kawanaka}, {Kawakubo}, {Kobayashi}, {Kohri}, {Krawczynski}, {Krizmanic},
  {Link}, {Maestro}, {Marrocchesi}, {Messineo}, {Mitchell}, {Miyake},
  {Moiseev}, {Mori}, {Mori}, {Motz}, {Munakata}, {Nakahira}, {Nishimura}, {de
  Nolfo}, {Okuno}, {Ormes}, {Ospina}, {Ozawa}, {Pacini}, {Palma}, {Papini},
  {Rauch}, {Ricciarini}, {Sakai}, {Sakamoto}, {Sasaki}, {Shimizu}, {Shiomi},
  {Sparvoli}, {Spillantini}, {Stolzi}, {Sugita}, {Suh}, {Sulaj}, {Takita},
  {Tamura}, {Terasawa}, {Torii}, {Tsunesada}, {Uchihori}, {Vannuccini},
  {Wefel}, {Yamaoka}, {Yanagita}, {Yoshida}, {Yoshida} \& {Calet
  Collaboration}}]{2020PhRvL.125y1102A}
\bibinfo{author}{{Adriani}, O.}, \bibinfo{author}{{Akaike}, Y.},
  \bibinfo{author}{{Asano}, K.} et~al. (\bibinfo{year}{2020}).
\newblock \bibinfo{title}{Direct measurement of the cosmic-ray carbon and
  oxygen spectra from 10 {GeV/n} to 2.2 {TeV/n} with the {Calorimetric Electron
  Telescope on the International Space Station}}.
\newblock {\it \bibinfo{journal}{\prl}\/},  {\it
  \bibinfo{volume}{125}\/}\bibinfo{issue}{(25)}, \bibinfo{pages}{251102}.
  \DOIprefix\doi{10.1103/PhysRevLett.125.251102}.
  \href{http://arxiv.org/abs/2012.10319}{\tt arXiv:2012.10319}.

\bibitem[{{Adriani} et~al.(2022{\natexlab{a}}){Adriani}, {Akaike}, {Asano},
  {Asaoka}, {Berti}, {Bigongiari}, {Binns}, {Bongi}, {Brogi}, {Bruno},
  {Buckley}, {Cannady}, {Castellini}, {Checchia}, {Cherry}, {Collazuol}, {de
  Nolfo}, {Ebisawa}, {Ficklin}, {Fuke}, {Gonzi}, {Guzik}, {Hams}, {Hibino},
  {Ichimura}, {Ioka}, {Ishizaki}, {Israel}, {Kasahara}, {Kataoka}, {Kataoka},
  {Katayose}, {Kato}, {Kawanaka}, {Kawakubo}, {Kobayashi}, {Kohri},
  {Krawczynski}, {Krizmanic}, {Maestro}, {Marrocchesi}, {Messineo}, {Mitchell},
  {Miyake}, {Moiseev}, {Mori}, {Mori}, {Motz}, {Munakata}, {Nakahira},
  {Nishimura}, {Okuno}, {Ormes}, {Ozawa}, {Pacini}, {Papini}, {Rauch},
  {Ricciarini}, {Sakai}, {Sakamoto}, {Sasaki}, {Shimizu}, {Shiomi},
  {Spillantini}, {Stolzi}, {Sugita}, {Sulaj}, {Takita}, {Tamura}, {Terasawa},
  {Torii}, {Tsunesada}, {Uchihori}, {Vannuccini}, {Wefel}, {Yamaoka},
  {Yanagita}, {Yoshida}, {Yoshida}, {Zober} \& {Calet
  Collaboration}}]{2022PhRvL.129y1103A}
\bibinfo{author}{{Adriani}, O.}, \bibinfo{author}{{Akaike}, Y.},
  \bibinfo{author}{{Asano}, K.} et~al. (\bibinfo{year}{2022}{\natexlab{a}}).
\newblock \bibinfo{title}{Cosmic-ray {Boron} flux measured from 8.4 {GeV/n} to
  3.8 {TeV/n} with the {Calorimetric Electron Telescope on the International
  Space Station}}.
\newblock {\it \bibinfo{journal}{\prl}\/},  {\it
  \bibinfo{volume}{129}\/}\bibinfo{issue}{(25)}, \bibinfo{pages}{251103}.
  \DOIprefix\doi{10.1103/PhysRevLett.129.251103}.
  \href{http://arxiv.org/abs/2212.07873}{\tt arXiv:2212.07873}.

\bibitem[{{Adriani} et~al.(2023){Adriani}, {Akaike}, {Asano}, {Asaoka},
  {Berti}, {Bigongiari}, {Binns}, {Bongi}, {Brogi}, {Bruno}, {Buckley},
  {Cannady}, {Castellini}, {Checchia}, {Cherry}, {Collazuol}, {de Nolfo},
  {Ebisawa}, {Ficklin}, {Fuke}, {Gonzi}, {Guzik}, {Hams}, {Hibino}, {Ichimura},
  {Ioka}, {Ishizaki}, {Israel}, {Kasahara}, {Kataoka}, {Kataoka}, {Katayose},
  {Kato}, {Kawanaka}, {Kawakubo}, {Kobayashi}, {Kohri}, {Krawczynski},
  {Krizmanic}, {Maestro}, {Marrocchesi}, {Messineo}, {Mitchell}, {Miyake},
  {Moiseev}, {Mori}, {Mori}, {Motz}, {Munakata}, {Nakahira}, {Nishimura},
  {Okuno}, {Ormes}, {Ozawa}, {Pacini}, {Papini}, {Rauch}, {Ricciarini},
  {Sakai}, {Sakamoto}, {Sasaki}, {Shimizu}, {Shiomi}, {Spillantini}, {Stolzi},
  {Sugita}, {Sulaj}, {Takita}, {Tamura}, {Terasawa}, {Torii}, {Tsunesada},
  {Uchihori}, {Vannuccini}, {Wefel}, {Yamaoka}, {Yanagita}, {Yoshida},
  {Yoshida}, {Zober} \& {Calet Collaboration}}]{2023PhRvL.130q1002A}
\bibinfo{author}{{Adriani}, O.}, \bibinfo{author}{{Akaike}, Y.},
  \bibinfo{author}{{Asano}, K.} et~al. (\bibinfo{year}{2023}).
\newblock \bibinfo{title}{Direct measurement of the cosmic-ray helium spectrum
  from 40 {GeV} to 250 {TeV with the Calorimetric Electron Telescope on the
  International Space Station}}.
\newblock {\it \bibinfo{journal}{\prl}\/},  {\it
  \bibinfo{volume}{130}\/}\bibinfo{issue}{(17)}, \bibinfo{pages}{171002}.
  \DOIprefix\doi{10.1103/PhysRevLett.130.171002}.
  \href{http://arxiv.org/abs/2304.14699}{\tt arXiv:2304.14699}.

\bibitem[{{Adriani} et~al.(2022{\natexlab{b}}){Adriani}, Akaike, Asano, Asaoka,
  Berti, Bigongiari, Binns, Bongi, Brogi, Bruno, Buckley, Cannady, Castellini,
  Checchia, Cherry, Collazuol, Ebisawa, Ficklin, Fuke, Gonzi, Guzik, Hams,
  Hibino, Ichimura, Ioka, Ishizaki, Israel, Kasahara, Kataoka, Kataoka,
  Katayose, Kato, Kawanaka, Kawakubo, Kobayashi, Kohri, Krawczynski, Krizmanic,
  Maestro, Marrocchesi, Messineo, Mitchell, Miyake, Moiseev, Mori, Mori, Motz,
  Munakata, Nakahira, Nishimura, de~Nolfo, Okuno, Ormes, Ozawa, Pacini, Papini,
  Rauch, Ricciarini, Sakai, Sakamoto, Sasaki, Shimizu, Shiomi, Spillantini,
  Stolzi, Sugita, Sulaj, Takita, Tamura, Terasawa, Torii, Tsunesada, Uchihori,
  Vannuccini, Wefel, Yamaoka, Yanagita, Yoshida, Yoshida \&
  Zober}]{AdrianiCalet2022}
\bibinfo{author}{{Adriani}, O.}, \bibinfo{author}{Akaike, Y.},
  \bibinfo{author}{Asano, K.} et~al. (\bibinfo{year}{2022}{\natexlab{b}}).
\newblock \bibinfo{title}{Observation of spectral structures in the flux of
  cosmic-ray protons from 50 {GeV} to 60 {TeV} with the {Calorimetric Electron
  Telescope on the International Space Station}}.
\newblock {\it \bibinfo{journal}{\prl}\/},  {\it \bibinfo{volume}{129}\/},
  \bibinfo{pages}{101102}. \URLprefix
  \url{https://link.aps.org/doi/10.1103/PhysRevLett.129.101102}.
  \DOIprefix\doi{10.1103/PhysRevLett.129.101102}.

\bibitem[{{Adriani} et~al.(2021){Adriani}, {Akaike}, {Asano}, {Asaoka},
  {Berti}, {Bigongiari}, {Binns}, {Bongi}, {Brogi}, {Bruno}, {Buckley},
  {Cannady}, {Castellini}, {Checchia}, {Cherry}, {Collazuol}, {Ebisawa},
  {Fuke}, {Gonzi}, {Guzik}, {Hams}, {Hibino}, {Ichimura}, {Ioka}, {Ishizaki},
  {Israel}, {Kasahara}, {Kataoka}, {Kataoka}, {Katayose}, {Kato}, {Kawanaka},
  {Kawakubo}, {Kobayashi}, {Kohri}, {Krawczynski}, {Krizmanic}, {Link},
  {Maestro}, {Marrocchesi}, {Messineo}, {Mitchell}, {Miyake}, {Moiseev},
  {Mori}, {Mori}, {Motz}, {Munakata}, {Nakahira}, {Nishimura}, {de Nolfo},
  {Okuno}, {Ormes}, {Ospina}, {Ozawa}, {Pacini}, {Papini}, {Rauch},
  {Ricciarini}, {Sakai}, {Sakamoto}, {Sasaki}, {Shimizu}, {Shiomi},
  {Spillantini}, {Stolzi}, {Sugita}, {Sulaj}, {Takita}, {Tamura}, {Terasawa},
  {Torii}, {Tsunesada}, {Uchihori}, {Vannuccini}, {Wefel}, {Yamaoka},
  {Yanagita}, {Yoshida}, {Yoshida} \& {Calet
  Collaboration}}]{2021PhRvL.126x1101A}
\bibinfo{author}{{Adriani}, O.}, \bibinfo{author}{{Akaike}, Y.},
  \bibinfo{author}{{Asano}, K.} et~al. (\bibinfo{year}{2021}).
\newblock \bibinfo{title}{Measurement of the iron spectrum in cosmic rays from
  10 {GeV/n} to 2.0 {TeV/n} with the {Calorimetric Electron Telescope on the
  International Space Station}}.
\newblock {\it \bibinfo{journal}{\prl}\/},  {\it
  \bibinfo{volume}{126}\/}\bibinfo{issue}{(24)}, \bibinfo{pages}{241101}.
  \DOIprefix\doi{10.1103/PhysRevLett.126.241101}.
  \href{http://arxiv.org/abs/2106.08036}{\tt arXiv:2106.08036}.

\bibitem[{{Adriani} et~al.(2011){Adriani}, {Barbarino}, {Bazilevskaya},
  {Bellotti}, {Boezio}, {Bogomolov}, {Bonechi}, {Bongi}, {Bonvicini},
  {Borisov}, {Bottai}, {Bruno}, {Cafagna}, {Campana}, {Carbone}, {Carlson},
  {Casolino}, {Castellini}, {Consiglio}, {De Pascale}, {De Santis}, {De
  Simone}, {Di Felice}, {Galper}, {Gillard}, {Grishantseva}, {Jerse},
  {Karelin}, {Koldashov}, {Krutkov}, {Kvashnin}, {Leonov}, {Malakhov},
  {Malvezzi}, {Marcelli}, {Mayorov}, {Menn}, {Mikhailov}, {Mocchiutti},
  {Monaco}, {Mori}, {Nikonov}, {Osteria}, {Palma}, {Papini}, {Pearce},
  {Picozza}, {Pizzolotto}, {Ricci}, {Ricciarini}, {Rossetto}, {Sarkar},
  {Simon}, {Sparvoli}, {Spillantini}, {Stozhkov}, {Vacchi}, {Vannuccini},
  {Vasilyev}, {Voronov}, {Yurkin}, {Wu}, {Zampa}, {Zampa} \&
  {Zverev}}]{2011Sci...332...69A}
\bibinfo{author}{{Adriani}, O.}, \bibinfo{author}{{Barbarino}, G.~C.},
  \bibinfo{author}{{Bazilevskaya}, G.~A.} et~al. (\bibinfo{year}{2011}).
\newblock \bibinfo{title}{{PAMELA} measurements of cosmic-ray proton and helium
  spectra}.
\newblock {\it \bibinfo{journal}{Science}\/},  {\it
  \bibinfo{volume}{332}\/}\bibinfo{issue}{(6025)}, \bibinfo{pages}{69}.
  \DOIprefix\doi{10.1126/science.1199172}.
  \href{http://arxiv.org/abs/1103.4055}{\tt arXiv:1103.4055}.

\bibitem[{{Aguilar} et~al.(2015){Aguilar}, {Aisa}, {Alpat}, {Alvino},
  {Ambrosi}, {Andeen}, {Arruda}, {Attig}, {Azzarello}, {Bachlechner}, {Barao},
  {Barrau}, {Barrin}, {Bartoloni}, {Basara}, {Battarbee}, {Battiston}, {Bazo},
  {Becker}, {Behlmann}, {Beischer}, {Berdugo}, {Bertucci}, {Bigongiari},
  {Bindi}, {Bizzaglia}, {Bizzarri}, {Boella}, {de Boer}, {Bollweg},
  {Bonnivard}, {Borgia}, {Borsini}, {Boschini}, {Bourquin}, {Burger}, {Cadoux},
  {Cai}, {Capell}, {Caroff}, {Casaus}, {Cascioli}, {Castellini}, {Cernuda},
  {Cerreta}, {Cervelli}, {Chae}, {Chang}, {Chen}, {Chen}, {Cheng}, {Chen},
  {Cheng}, {Chou}, {Choumilov}, {Choutko}, {Chung}, {Clark}, {Clavero},
  {Coignet}, {Consolandi}, {Contin}, {Corti}, {Gil}, {Coste}, {Creus},
  {Crispoltoni}, {Cui}, {Dai}, {Delgado}, {Della Torre}, {Demirk{\"o}z},
  {Derome}, {Di Falco}, {Di Masso}, {Dimiccoli}, {D{\'\i}az}, {von Doetinchem},
  {Donnini}, {Du}, {Duranti}, {D'Urso}, {Eline}, {Eppling}, {Eronen}, {Fan},
  {Farnesini}, {Feng}, {Fiandrini}, {Fiasson}, {Finch}, {Fisher},
  {Galaktionov}, {Gallucci}, {Garc{\'\i}a}, {Garc{\'\i}a-L{\'o}pez},
  {Gargiulo}, {Gast}, {Gebauer} et~al.}]{2015PhRvL.114q1103A}
\bibinfo{author}{{Aguilar}, M.}, \bibinfo{author}{{Aisa}, D.},
  \bibinfo{author}{{Alpat}, B.} et~al. (\bibinfo{year}{2015}).
\newblock \bibinfo{title}{Precision measurement of the proton flux in primary
  cosmic rays from rigidity 1 {GV} to 1.8 {TV} with the {Alpha Magnetic
  Spectrometer on the International Space Station}}.
\newblock {\it \bibinfo{journal}{\prl}\/},  {\it
  \bibinfo{volume}{114}\/}\bibinfo{issue}{(17)}, \bibinfo{pages}{171103}.
  \DOIprefix\doi{10.1103/PhysRevLett.114.171103}.

\bibitem[{{Aguilar} et~al.(2021{\natexlab{a}}){Aguilar}, {Ali Cavasonza},
  {Ambrosi}, {Arruda}, {Attig}, {Barao}, {Barrin}, {Bartoloni},
  {Ba{\c{s}}e{\u{g}}mez-du Pree}, {Bates}, {Battiston}, {Behlmann}, {Beischer},
  {Berdugo}, {Bertucci}, {Bindi}, {de Boer}, {Bollweg}, {Borgia}, {Boschini},
  {Bourquin}, {Bueno}, {Burger}, {Burger}, {Burmeister}, {Cai}, {Capell},
  {Casaus}, {Castellini}, {Cervelli}, {Chang}, {Chen}, {Chen}, {Chen}, {Cheng},
  {Chou}, {Chouridou}, {Choutko}, {Chung}, {Clark}, {Coignet}, {Consolandi},
  {Contin}, {Corti}, {Cui}, {Dadzie}, {Dai}, {Delgado}, {Della Torre},
  {Demirk{\"o}z}, {Derome}, {Di Falco}, {Di Felice}, {D{\'\i}az}, {Dimiccoli},
  {von Doetinchem}, {Dong}, {Donnini}, {Duranti}, {Egorov}, {Eline}, {Feng},
  {Fiandrini}, {Fisher}, {Formato}, {Freeman}, {Galaktionov}, {G{\'a}mez},
  {Garc{\'\i}a-L{\'o}pez}, {Gargiulo}, {Gast}, {Gebauer}, {Gervasi},
  {Giovacchini}, {G{\'o}mez-Coral}, {Gong}, {Goy}, {Grabski}, {Grandi},
  {Graziani}, {Guo}, {Haino}, {Han}, {Hashmani}, {He}, {Heber}, {Hsieh}, {Hu},
  {Huang}, {Hungerford}, {Incagli}, {Jang}, {Jia}, {Jinchi}, {Kanishev},
  {Khiali}, {Kim}, {Kirn}, {Konyushikhin} et~al.}]{Aguilar2021}
\bibinfo{author}{{Aguilar}, M.}, \bibinfo{author}{{Ali Cavasonza}, L.},
  \bibinfo{author}{{Ambrosi}, G.} et~al. (\bibinfo{year}{2021}{\natexlab{a}}).
\newblock \bibinfo{title}{{The Alpha Magnetic Spectrometer (AMS) on the
  International Space Station: Part II} -- results from the first seven years}.
\newblock {\it \bibinfo{journal}{\physrep}\/},  {\it \bibinfo{volume}{894}\/},
  \bibinfo{pages}{1--116}. \DOIprefix\doi{10.1016/j.physrep.2020.09.003}.

\bibitem[{{Aguilar} et~al.(2021{\natexlab{b}}){Aguilar}, {Cavasonza}, {Allen},
  {Alpat}, {Ambrosi}, {Arruda}, {Attig}, {Barao}, {Barrin}, {Bartoloni},
  {Ba{\c{s}}e{\v{g}}mez-du Pree}, {Battiston}, {Behlmann}, {Beischer},
  {Berdugo}, {Bertucci}, {Bindi}, {de Boer}, {Bollweg}, {Borgia}, {Boschini},
  {Bourquin}, {Bueno}, {Burger}, {Burger}, {Burmeister}, {Cai}, {Capell},
  {Casaus}, {Castellini}, {Cervelli}, {Chang}, {Chen}, {Chen}, {Chen}, {Chen},
  {Cheng}, {Chou}, {Chouridou}, {Choutko}, {Chung}, {Clark}, {Coignet},
  {Consolandi}, {Contin}, {Corti}, {Cui}, {Dadzie}, {Delgado}, {Della Torre},
  {Demirk{\"o}z}, {Derome}, {Di Falco}, {Di Felice}, {D{\'\i}az}, {Dimiccoli},
  {von Doetinchem}, {Dong}, {Donnini}, {Duranti}, {Egorov}, {Eline}, {Feng},
  {Fiandrini}, {Fisher}, {Formato}, {Freeman}, {Galaktionov}, {G{\'a}mez},
  {Garc{\'\i}a-L{\'o}pez}, {Gargiulo}, {Gast}, {Gervasi}, {Giovacchini},
  {G{\'o}mez-Coral}, {Gong}, {Goy}, {Grabski}, {Grandi}, {Graziani}, {Haino},
  {Han}, {Hashmani}, {He}, {Heber}, {Hsieh}, {Hu}, {Incagli}, {Jang}, {Jia},
  {Jinchi}, {Kanishev}, {Khiali}, {Kim}, {Kirn}, {Konyushikhin}, {Kounina},
  {Kounine}, {Koutsenko} et~al.}]{2021PhRvL.126d1104A}
\bibinfo{author}{{Aguilar}, M.}, \bibinfo{author}{{Cavasonza}, L.~A.},
  \bibinfo{author}{{Allen}, M.~S.} et~al. (\bibinfo{year}{2021}{\natexlab{b}}).
\newblock \bibinfo{title}{Properties of iron primary cosmic rays: Results from
  the {Alpha Magnetic Spectrometer}}.
\newblock {\it \bibinfo{journal}{\prl}\/},  {\it
  \bibinfo{volume}{126}\/}\bibinfo{issue}{(4)}, \bibinfo{pages}{041104}.
  \DOIprefix\doi{10.1103/PhysRevLett.126.041104}.

\bibitem[{{Ahn} et~al.(2009){Ahn}, {Allison}, {Bagliesi}, {Barbier}, {Beatty},
  {Bigongiari}, {Brandt}, {Childers}, {Conklin}, {Coutu}, {Du Vernois},
  {Ganel}, {Han}, {Jeon}, {Kim}, {Lee}, {Maestro}, {Malinine}, {Marrocchesi},
  {Minnick}, {Mognet}, {Nam}, {Nutter}, {Park}, {Park}, {Seo}, {Sina},
  {Walpole}, {Wu}, {Yang}, {Yoon}, {Zei} \& {Zinn}}]{2009ApJ...707..593A}
\bibinfo{author}{{Ahn}, H.~S.}, \bibinfo{author}{{Allison}, P.},
  \bibinfo{author}{{Bagliesi}, M.~G.} et~al. (\bibinfo{year}{2009}).
\newblock \bibinfo{title}{Energy spectra of cosmic-ray nuclei at high
  energies}.
\newblock {\it \bibinfo{journal}{\apj}\/},  {\it
  \bibinfo{volume}{707}\/}\bibinfo{issue}{(1)}, \bibinfo{pages}{593--603}.
  \DOIprefix\doi{10.1088/0004-637X/707/1/593}.
  \href{http://arxiv.org/abs/0911.1889}{\tt arXiv:0911.1889}.

\bibitem[{{Alemanno} et~al.(2021){Alemanno}, {An}, {Azzarello}, {Barbato},
  {Bernardini}, {Bi}, {Cai}, {Catanzani}, {Chang}, {Chen}, {Chen}, {Chen},
  {Cui}, {Cui}, {Cui}, {Dai}, {D'Amone}, {de Benedittis}, {de Mitri}, {de
  Palma}, {Deliyergiyev}, {di Santo}, {Dong}, {Dong}, {Donvito}, {Droz},
  {Duan}, {Duan}, {D'Urso}, {Fan}, {Fan}, {Fang}, {Fang}, {Feng}, {Feng},
  {Fusco}, {Gao}, {Gargano}, {Gong}, {Gong}, {Guo}, {Guo}, {Guo}, {Han}, {Hu},
  {Huang}, {Huang}, {Huang}, {Ionica}, {Jiang}, {Kong}, {Kotenko}, {Kyratzis},
  {Lei}, {Li}, {Li}, {Li}, {Li}, {Liang}, {Liu}, {Liu}, {Liu}, {Liu}, {Liu},
  {Liu}, {Loparco}, {Luo}, {Ma}, {Ma}, {Ma}, {Ma}, {Marsella}, {Mazziotta},
  {Mo}, {Niu}, {Pan}, {Parenti}, {Peng}, {Peng}, {Perrina}, {Qiao}, {Rao},
  {Ruina}, {Salinas}, {Shang}, {Shen}, {Shen}, {Shen}, {Silveri}, {Song},
  {Stolpovskiy}, {Su}, {Su}, {Sun}, {Surdo}, {Teng}, {Tykhonov}, {Wang}, {Wang}
  et~al.}]{2021PhRvL.126t1102A}
\bibinfo{author}{{Alemanno}, F.}, \bibinfo{author}{{An}, Q.},
  \bibinfo{author}{{Azzarello}, P.} et~al. (\bibinfo{year}{2021}).
\newblock \bibinfo{title}{Measurement of the cosmic ray helium energy spectrum
  from 70 {GeV} to 80 {TeV} with the {DAMPE} space mission}.
\newblock {\it \bibinfo{journal}{\prl}\/},  {\it
  \bibinfo{volume}{126}\/}\bibinfo{issue}{(20)}, \bibinfo{pages}{201102}.
  \DOIprefix\doi{10.1103/PhysRevLett.126.201102}.
  \href{http://arxiv.org/abs/2105.09073}{\tt arXiv:2105.09073}.

\bibitem[{{Amenomori} et~al.(2017){Amenomori}, {Bi}, {Chen}, {Chen}, {Chen},
  {Cui}, {Danzengluobu}, {Ding}, {Feng}, {Feng}, {Feng}, {Gou}, {Guo}, {He},
  {He}, {Hibino}, {Hotta}, {Hu}, {Hu}, {Huang}, {Jia}, {Jiang}, {Kajino},
  {Kasahara}, {Katayose}, {Kato}, {Kawata}, {Kozai}, {Labaciren}, {Le}, {Li},
  {Li}, {Li}, {Liu}, {Liu}, {Liu}, {Lu}, {Meng}, {Miyazaki}, {Mizutani},
  {Munakata}, {Nakajima}, {Nakamura}, {Nanjo}, {Nishizawa}, {Niwa}, {Ohnishi},
  {Ohta}, {Ozawa}, {Qian}, {Qu}, {Saito}, {Saito}, {Sakata}, {Sako}, {Shao},
  {Shibata}, {Shiomi}, {Shirai}, {Sugimoto}, {Takita}, {Tan}, {Tateyama},
  {Torii}, {Tsuchiya}, {Udo}, {Wang}, {Wu}, {Xue}, {Yamamoto}, {Yamauchi},
  {Yang}, {Yuan}, {Yuda}, {Zhai}, {Zhang}, {Zhang}, {Zhang}, {Zhang}, {Zhang},
  {Zhang}, {Zhaxisangzhu}, {Zhou} \& {Tibet AS{\ensuremath{\gamma}}
  Collaboration}}]{Amenomori2017}
\bibinfo{author}{{Amenomori}, M.}, \bibinfo{author}{{Bi}, X.~J.},
  \bibinfo{author}{{Chen}, D.} et~al. (\bibinfo{year}{2017}).
\newblock \bibinfo{title}{Northern sky galactic cosmic ray anisotropy between
  10 and 1000 {TeV} with the {Tibet Air Shower Array}}.
\newblock {\it \bibinfo{journal}{\apj}\/},  {\it
  \bibinfo{volume}{836}\/}\bibinfo{issue}{(2)}, \bibinfo{pages}{153}.
  \DOIprefix\doi{10.3847/1538-4357/836/2/153}.
  \href{http://arxiv.org/abs/1701.07144}{\tt arXiv:1701.07144}.

\bibitem[{An et~al.(2019)An, Asfandiyarov, Azzarello, Bernardini, Bi, Cai,
  Chang, Chen, Chen, Chen, Chen, Cui, Cui, Dai, D{\textquoteright}Amone,
  De~Benedittis, De~Mitri, Di~Santo, Ding, Dong, Dong, Dong, Donvito, Droz,
  Duan, Duan, D{\textquoteright}Urso, Fan, Fan, Fang, Feng, Feng, Fusco, Gallo,
  Gan, Gao, Gargano, Gong, Gong, Guo, Guo, Guo, Han, Hu, Huang, Huang, Huang,
  Ionica, Jiang, Jin, Kong, Lei, Li, Li, Li, Li, Li, Liang, Liang, Liao, Liu,
  Liu, Liu, Liu, Liu, Liu, Loparco, Luo, Ma, Ma, Ma, Ma, Ma, Marsella,
  Mazziotta, Mo, Niu, Pan, Peng, Peng, Qiao, Rao, Salinas, Shang, Shen, Shen,
  Shen, Song, Su, Su, Sun, Surdo, Teng, Tykhonov, Vitillo, Wang, Wang, Wang,
  Wang et~al.}]{Dampe2019a}
\bibinfo{author}{An, Q.}, \bibinfo{author}{Asfandiyarov, R.},
  \bibinfo{author}{Azzarello, P.} et~al. (\bibinfo{year}{2019}).
\newblock \bibinfo{title}{Measurement of the cosmic ray proton spectrum from 40
  {GeV} to 100 {TeV} with the {DAMPE} satellite}.
\newblock {\it \bibinfo{journal}{\sa}\/},  {\it
  \bibinfo{volume}{5}\/}\bibinfo{issue}{(9)}. \URLprefix
  \url{https://www.science.org/doi/10.1126/sciadv.aax3793}.
  \DOIprefix\doi{10.1126/sciadv.aax3793}.

\bibitem[{{Becker Tjus} \& {Merten}(2020)}]{BeckerTjus2020}
\bibinfo{author}{{Becker Tjus}, J.},  \& \bibinfo{author}{{Merten}, L.}
  (\bibinfo{year}{2020}).
\newblock \bibinfo{title}{{Closing in on the origin of Galactic cosmic rays
  using multimessenger information}}.
\newblock {\it \bibinfo{journal}{\physrep}\/},  {\it \bibinfo{volume}{872}\/},
  \bibinfo{pages}{1--98}. \DOIprefix\doi{10.1016/j.physrep.2020.05.002}.
  \href{http://arxiv.org/abs/2002.00964}{\tt arXiv:2002.00964}.

\bibitem[{{Blandford} \& {Ostriker}(1978)}]{BlandOst78}
\bibinfo{author}{{Blandford}, R.~D.},  \& \bibinfo{author}{{Ostriker}, J.~P.}
  (\bibinfo{year}{1978}).
\newblock \bibinfo{title}{{Particle acceleration by astrophysical shocks}}.
\newblock {\it \bibinfo{journal}{\apjl}\/},  {\it \bibinfo{volume}{221}\/},
  \bibinfo{pages}{L29--L32}. \DOIprefix\doi{10.1086/182658}.

\bibitem[{{Blasi} et~al.(2022){Blasi}, {Amato}, {Aloisio}, {Dundovic}, {Evoli},
  {Gaggero}, {Grasso} \& {Morlino}}]{Blasi2022}
\bibinfo{author}{{Blasi}, P.}, \bibinfo{author}{{Amato}, E.},
  \bibinfo{author}{{Aloisio}, R.} et~al. (\bibinfo{year}{2022}).
\newblock \bibinfo{title}{A modern approach to cosmic ray transport in the
  {Galaxy}}.
\newblock {\it \bibinfo{journal}{\memsai}\/},  {\it
  \bibinfo{volume}{93}\/}\bibinfo{issue}{(2-3)}, \bibinfo{pages}{168}.
  \DOIprefix\doi{10.36116/MEMSAIT-93N2-3.2022.22}.

\bibitem[{{Boschini} et~al.(2020){Boschini}, {Della Torre}, {Gervasi},
  {Grandi}, {J{\'o}hannesson}, {La Vacca}, {Masi}, {Moskalenko}, {Pensotti},
  {Porter}, {Quadrani}, {Rancoita}, {Rozza} \& {Tacconi}}]{2020ApJS..250...27B}
\bibinfo{author}{{Boschini}, M.~J.}, \bibinfo{author}{{Della Torre}, S.},
  \bibinfo{author}{{Gervasi}, M.} et~al. (\bibinfo{year}{2020}).
\newblock \bibinfo{title}{Inference of the local interstellar spectra of
  cosmic-ray nuclei {$Z {\leq} 28$} with the {GALPROP-HELMOD} framework}.
\newblock {\it \bibinfo{journal}{\apjs}\/},  {\it
  \bibinfo{volume}{250}\/}\bibinfo{issue}{(2)}, \bibinfo{pages}{27}.
  \DOIprefix\doi{10.3847/1538-4365/aba901}.
  \href{http://arxiv.org/abs/2006.01337}{\tt arXiv:2006.01337}.

\bibitem[{{Boschini} et~al.(2021){Boschini}, {Della Torre}, {Gervasi},
  {Grandi}, {J{\'o}hannesson}, {La Vacca}, {Masi}, {Moskalenko}, {Pensotti},
  {Porter}, {Quadrani}, {Rancoita}, {Rozza} \& {Tacconi}}]{2021ApJ...913....5B}
\bibinfo{author}{{Boschini}, M.~J.}, \bibinfo{author}{{Della Torre}, S.},
  \bibinfo{author}{{Gervasi}, M.} et~al. (\bibinfo{year}{2021}).
\newblock \bibinfo{title}{The discovery of a low-energy excess in cosmic-ray
  iron: Evidence of the past supernova activity in the {Local Bubble}}.
\newblock {\it \bibinfo{journal}{\apj}\/},  {\it
  \bibinfo{volume}{913}\/}\bibinfo{issue}{(1)}, \bibinfo{pages}{5}.
  \DOIprefix\doi{10.3847/1538-4357/abf11c}.
  \href{http://arxiv.org/abs/2101.12735}{\tt arXiv:2101.12735}.

\bibitem[{{Burlaga} et~al.(2019){Burlaga}, {Ness}, {Berdichevsky}, {Park},
  {Jian}, {Szabo}, {Stone} \& {Richardson}}]{Burlaga2019}
\bibinfo{author}{{Burlaga}, L.~F.}, \bibinfo{author}{{Ness}, N.~F.},
  \bibinfo{author}{{Berdichevsky}, D.~B.} et~al. (\bibinfo{year}{2019}).
\newblock \bibinfo{title}{{Magnetic field and particle measurements made by
  Voyager 2 at and near the heliopause}}.
\newblock {\it \bibinfo{journal}{Nature Astronomy}\/},  {\it
  \bibinfo{volume}{3}\/}, \bibinfo{pages}{1007--1012}.
  \DOIprefix\doi{10.1038/s41550-019-0920-y}.

\bibitem[{Chakraborty et~al.(2023)Chakraborty, Ahmad, Chandra, Dugad, Goswami,
  Gupta, Hariharan, Hayashi, Jagadeesan, Jain et~al.}]{Chakraborty2023}
\bibinfo{author}{Chakraborty, M.}, \bibinfo{author}{Ahmad, S.},
  \bibinfo{author}{Chandra, A.} et~al. (\bibinfo{year}{2023}).
\newblock \bibinfo{title}{Small-scale cosmic ray anisotropy observed by the
  {GRAPES-3} experiment at {TeV} energies}.
\newblock {\it \bibinfo{journal}{arXiv:2310.15489}\/}, .

\bibitem[{Chen et~al.(2020)Chen, Bale, Bonnell, Borovikov, Bowen, Burgess,
  Case, Chandran, de~Wit, Goetz, Harvey, Kasper, Klein, Korreck, Larson, Livi,
  MacDowall, Malaspina, Mallet, McManus, Moncuquet, Pulupa, Stevens \&
  Whittlesey}]{Chen_2020}
\bibinfo{author}{Chen, C. H.~K.}, \bibinfo{author}{Bale, S.~D.},
  \bibinfo{author}{Bonnell, J.~W.} et~al. (\bibinfo{year}{2020}).
\newblock \bibinfo{title}{The evolution and role of solar wind turbulence in
  the inner heliosphere}.
\newblock {\it \bibinfo{journal}{\apjs}\/},  {\it
  \bibinfo{volume}{246}\/}\bibinfo{issue}{(2)}, \bibinfo{pages}{53}. \URLprefix
  \url{https://doi.org/10.3847/1538-4365/ab60a3}.
  \DOIprefix\doi{10.3847/1538-4365/ab60a3}.

\bibitem[{{Chernyshov} et~al.(2022){Chernyshov}, {Dogiel}, {Ivlev}, {Erlykin}
  \& {Kiselev}}]{Chernyshov2022}
\bibinfo{author}{{Chernyshov}, D.~O.}, \bibinfo{author}{{Dogiel}, V.~A.},
  \bibinfo{author}{{Ivlev}, A.~V.} et~al. (\bibinfo{year}{2022}).
\newblock \bibinfo{title}{Formation of the cosmic-ray halo: The role of
  nonlinear {Landau} damping}.
\newblock {\it \bibinfo{journal}{\apj}\/},  {\it
  \bibinfo{volume}{937}\/}\bibinfo{issue}{(2)}, \bibinfo{pages}{107}.
  \DOIprefix\doi{10.3847/1538-4357/ac8f42}.
  \href{http://arxiv.org/abs/2209.12302}{\tt arXiv:2209.12302}.

\bibitem[{{Chernyshov} et~al.(2023){Chernyshov}, {Ivlev} \&
  {Dogiel}}]{ChernyshovIvlevDogiel2023}
\bibinfo{author}{{Chernyshov}, D.~O.}, \bibinfo{author}{{Ivlev}, A.~V.},  \&
  \bibinfo{author}{{Dogiel}, V.~A.} (\bibinfo{year}{2023}).
\newblock \bibinfo{title}{Secondary cosmic-ray nuclei in the model of
  {Galactic} halo with nonlinear {Landau} damping}.
\newblock {\it \bibinfo{journal}{arXiv e-prints}\/},  (p.
  \bibinfo{pages}{arXiv:2309.04772}).
  \DOIprefix\doi{10.48550/arXiv.2309.04772}.
  \href{http://arxiv.org/abs/2309.04772}{\tt arXiv:2309.04772}.

\bibitem[{Choi et~al.(2022)Choi, Seo, Aggarwal, Amare, Angelaszek, Bowman,
  Chen, Copley, Derome, Eraud et~al.}]{ISS-Cream2022}
\bibinfo{author}{Choi, G.}, \bibinfo{author}{Seo, E.},
  \bibinfo{author}{Aggarwal, S.} et~al. (\bibinfo{year}{2022}).
\newblock \bibinfo{title}{Measurement of high-energy cosmic-ray proton spectrum
  from the {ISS-CREAM} experiment}.
\newblock {\it \bibinfo{journal}{\apj}\/},  {\it
  \bibinfo{volume}{940}\/}\bibinfo{issue}{(2)}, \bibinfo{pages}{107}.

\bibitem[{Gabici(2023)}]{Gabici:2023/V}
\bibinfo{author}{Gabici, S.} (\bibinfo{year}{2023}).
\newblock \bibinfo{title}{{Rapporteur Talk: CRD}}.
\newblock In {\it \bibinfo{booktitle}{Proceedings of 38th International Cosmic
  Ray Conference {\textemdash} PoS(ICRC2023)}\/} (p. \bibinfo{pages}{030}).
\newblock volume \bibinfo{volume}{444}.
\newblock \DOIprefix\doi{10.22323/1.444.0030}.

\bibitem[{{Gaia Collaboration}(2020)}]{GaiaCollaboration2020}
\bibinfo{author}{{Gaia Collaboration}} (\bibinfo{year}{2020}).
\newblock \bibinfo{title}{{VizieR Online Data Catalog: Gaia EDR3 (Gaia
  Collaboration, 2020)}}.
\newblock {\it \bibinfo{journal}{VizieR Online Data Catalog}\/},  (p.
  \bibinfo{pages}{I/350}).

\bibitem[{{Grebenyuk} et~al.(2019){Grebenyuk}, {Karmanov}, {Kovalev},
  {Kudryashov}, {Kurganov}, {Panov}, {Podorozhny}, {Tkachenko}, {Tkachev},
  {Turundaevskiy}, {Vasiliev} \& {Voronin}}]{2019AdSpR..64.2546G}
\bibinfo{author}{{Grebenyuk}, V.}, \bibinfo{author}{{Karmanov}, D.},
  \bibinfo{author}{{Kovalev}, I.} et~al. (\bibinfo{year}{2019}).
\newblock \bibinfo{title}{Energy spectra of abundant cosmic-ray nuclei in the
  {NUCLEON} experiment}.
\newblock {\it \bibinfo{journal}{Advances in Space Research}\/},  {\it
  \bibinfo{volume}{64}\/}\bibinfo{issue}{(12)}, \bibinfo{pages}{2546--2558}.
  \DOIprefix\doi{10.1016/j.asr.2019.10.004}.

\bibitem[{{Karmanov} et~al.(2020{\natexlab{a}}){Karmanov}, {Kovalev},
  {Kudryashov}, {Kurganov}, {Panov}, {Podorozhny}, {Turundaevskiy} \&
  {Vasiliev}}]{2020PhLB..81135851K}
\bibinfo{author}{{Karmanov}, D.}, \bibinfo{author}{{Kovalev}, I.},
  \bibinfo{author}{{Kudryashov}, I.} et~al.
  (\bibinfo{year}{2020}{\natexlab{a}}).
\newblock \bibinfo{title}{Spectra of cosmic ray carbon and oxygen nuclei
  according to the {NUCLEON} experiment}.
\newblock {\it \bibinfo{journal}{Physics Letters B}\/},  {\it
  \bibinfo{volume}{811}\/}, \bibinfo{pages}{135851}.
  \DOIprefix\doi{10.1016/j.physletb.2020.135851}.

\bibitem[{{Karmanov} et~al.(2020{\natexlab{b}}){Karmanov}, {Kovalev},
  {Kudryashov}, {Kurganov}, {Panov}, {Podorozhny}, {Turundaevskiy} \&
  {Vasiliev}}]{2020JETPL.111..363K}
\bibinfo{author}{{Karmanov}, D.~E.}, \bibinfo{author}{{Kovalev}, I.~M.},
  \bibinfo{author}{{Kudryashov}, I.~A.} et~al.
  (\bibinfo{year}{2020}{\natexlab{b}}).
\newblock \bibinfo{title}{Spectra of protons and alpha particles and their
  comparison in the {NUCLEON} experiment data}.
\newblock {\it \bibinfo{journal}{Soviet Journal of Experimental and Theoretical
  Physics Letters}\/},  {\it \bibinfo{volume}{111}\/}\bibinfo{issue}{(7)},
  \bibinfo{pages}{363--367}. \DOIprefix\doi{10.1134/S002136402007005X}.

\bibitem[{{Kulsrud} \& {Pearce}(1969)}]{KulsrNeutr69}
\bibinfo{author}{{Kulsrud}, R.},  \& \bibinfo{author}{{Pearce}, W.~P.}
  (\bibinfo{year}{1969}).
\newblock \bibinfo{title}{The effect of wave-particle interactions on the
  propagation of cosmic rays}.
\newblock {\it \bibinfo{journal}{\apj}\/},  {\it \bibinfo{volume}{156}\/},
  \bibinfo{pages}{445--469}. \DOIprefix\doi{10.1086/149981}.

\bibitem[{Lallement et~al.(2005)Lallement, Qu{\'e}merais, Bertaux, Ferron,
  Koutroumpa \& Pellinen}]{Lallement2005}
\bibinfo{author}{Lallement, R.}, \bibinfo{author}{Qu{\'e}merais, E.},
  \bibinfo{author}{Bertaux, J.-L.} et~al. (\bibinfo{year}{2005}).
\newblock \bibinfo{title}{Deflection of the interstellar neutral hydrogen flow
  across the heliospheric interface}.
\newblock {\it \bibinfo{journal}{Science}\/},  {\it
  \bibinfo{volume}{307}\/}\bibinfo{issue}{(5714)}, \bibinfo{pages}{1447--1449}.

\bibitem[{{Lee} \& {V{\"o}lk}(1973)}]{Lee1973}
\bibinfo{author}{{Lee}, M.~A.},  \& \bibinfo{author}{{V{\"o}lk}, H.~J.}
  (\bibinfo{year}{1973}).
\newblock \bibinfo{title}{Damping and non-linear wave-particle interactions of
  {Alfv{\'e}n}-waves in the solar wind}.
\newblock {\it \bibinfo{journal}{\apss}\/},  {\it
  \bibinfo{volume}{24}\/}\bibinfo{issue}{(1)}, \bibinfo{pages}{31--49}.
  \DOIprefix\doi{10.1007/BF00648673}.

\bibitem[{Malkov(2017)}]{malkov2017exact}
\bibinfo{author}{Malkov, M.~A.} (\bibinfo{year}{2017}).
\newblock \bibinfo{title}{Exact solution of the {F}okker-{P}lanck equation for
  isotropic scattering}.
\newblock {\it \bibinfo{journal}{\prd}\/},  {\it
  \bibinfo{volume}{95}\/}\bibinfo{issue}{(2)}, \bibinfo{pages}{023007}.

\bibitem[{{Malkov} et~al.(2013){Malkov}, {Diamond}, {Sagdeev}, {Aharonian} \&
  {Moskalenko}}]{MetalEsc13}
\bibinfo{author}{{Malkov}, M.~A.}, \bibinfo{author}{{Diamond}, P.~H.},
  \bibinfo{author}{{Sagdeev}, R.~Z.} et~al. (\bibinfo{year}{2013}).
\newblock \bibinfo{title}{Analytic solution for self-regulated collective
  escape of cosmic rays from their acceleration sites}.
\newblock {\it \bibinfo{journal}{\apj}\/},  {\it \bibinfo{volume}{768}\/},
  \bibinfo{pages}{73}. \DOIprefix\doi{10.1088/0004-637X/768/1/73}.
  \href{http://arxiv.org/abs/1207.4728}{\tt arXiv:1207.4728}.

\bibitem[{{Malkov} \& {Moskalenko}(2021)}]{MalkMosk_2021}
\bibinfo{author}{{Malkov}, M.~A.},  \& \bibinfo{author}{{Moskalenko}, I.~V.}
  (\bibinfo{year}{2021}).
\newblock \bibinfo{title}{The {TeV} cosmic-ray bump: A message from the
  {Epsilon Indi} or {Epsilon Eridani} star? ({Paper~I})}.
\newblock {\it \bibinfo{journal}{\apj}\/},  {\it
  \bibinfo{volume}{911}\/}\bibinfo{issue}{(2)}, \bibinfo{pages}{151}.
  \URLprefix \url{https://doi.org/10.3847/1538-4357/abe855}.
  \DOIprefix\doi{10.3847/1538-4357/abe855}.

\bibitem[{{Malkov} \& {Moskalenko}(2022)}]{MalkMosk2022}
\bibinfo{author}{{Malkov}, M.~A.},  \& \bibinfo{author}{{Moskalenko}, I.~V.}
  (\bibinfo{year}{2022}).
\newblock \bibinfo{title}{On the origin of observed cosmic-ray spectrum below
  100 {TV} ({Paper~II})}.
\newblock {\it \bibinfo{journal}{\apj}\/},  {\it
  \bibinfo{volume}{933}\/}\bibinfo{issue}{(1)}, \bibinfo{pages}{78}.
  \DOIprefix\doi{10.3847/1538-4357/ac7049}.
  \href{http://arxiv.org/abs/2105.04630}{\tt arXiv:2105.04630}.

\bibitem[{{Panov} et~al.(2009){Panov}, {Adams}, {Ahn}, {Bashinzhagyan},
  {Watts}, {Wefel}, {Wu}, {Ganel}, {Guzik}, {Zatsepin}, {Isbert}, {Kim},
  {Christl}, {Kouznetsov}, {Panasyuk}, {Seo}, {Sokolskaya}, {Chang}, {Schmidt}
  \& {Fazely}}]{2009BRASP..73..564P}
\bibinfo{author}{{Panov}, A.~D.}, \bibinfo{author}{{Adams}, J.~H.},
  \bibinfo{author}{{Ahn}, H.~S.} et~al. (\bibinfo{year}{2009}).
\newblock \bibinfo{title}{Energy spectra of abundant nuclei of primary cosmic
  rays from the data of {ATIC-2} experiment: Final results}.
\newblock {\it \bibinfo{journal}{Bull.\ Russian Acad.\ Sci., Physics}\/},  {\it
  \bibinfo{volume}{73}\/}\bibinfo{issue}{(5)}, \bibinfo{pages}{564--567}.
  \DOIprefix\doi{10.3103/S1062873809050098}.
  \href{http://arxiv.org/abs/1101.3246}{\tt arXiv:1101.3246}.

\bibitem[{{Qiao} et~al.(2023){Qiao}, {Guo}, {Liu} \&
  {Bi}}]{2023ApJ...956...75Q}
\bibinfo{author}{{Qiao}, B.-Q.}, \bibinfo{author}{{Guo}, Y.-Q.},
  \bibinfo{author}{{Liu}, W.} et~al. (\bibinfo{year}{2023}).
\newblock \bibinfo{title}{Nearby {SNR:} a possible common origin of
  multi-messenger anomalies in the spectra, ratios, and anisotropy of cosmic
  rays}.
\newblock {\it \bibinfo{journal}{\apj}\/},  {\it
  \bibinfo{volume}{956}\/}\bibinfo{issue}{(2)}, \bibinfo{pages}{75}.
  \DOIprefix\doi{10.3847/1538-4357/acf453}.
  \href{http://arxiv.org/abs/2212.05641}{\tt arXiv:2212.05641}.

\bibitem[{{Riley} et~al.(2019){Riley}, {Strigari}, {Porter}, {Blandford},
  {Murgia}, {Kerr} \& {J{\'o}hannesson}}]{Riley2019}
\bibinfo{author}{{Riley}, A.~H.}, \bibinfo{author}{{Strigari}, L.~E.},
  \bibinfo{author}{{Porter}, T.~A.} et~al. (\bibinfo{year}{2019}).
\newblock \bibinfo{title}{Possible detection of gamma-rays from {Epsilon
  Eridani}}.
\newblock {\it \bibinfo{journal}{\apj}\/},  {\it
  \bibinfo{volume}{878}\/}\bibinfo{issue}{(1)}, \bibinfo{pages}{8}.
  \DOIprefix\doi{10.3847/1538-4357/ab1a3c}.
  \href{http://arxiv.org/abs/1810.04194}{\tt arXiv:1810.04194}.

\bibitem[{{Sagdeev} \& {Galeev}(1969)}]{SagdGal69}
\bibinfo{author}{{Sagdeev}, R.~Z.},  \& \bibinfo{author}{{Galeev}, A.~A.}
  (\bibinfo{year}{1969}).
\newblock {\it \bibinfo{title}{{Nonlinear Plasma Theory}}\/}.
\newblock \bibinfo{publisher}{W.A. Benjamin Inc. New York, New York}.

\bibitem[{{Skilling}(1975{\natexlab{a}})}]{Skill75a}
\bibinfo{author}{{Skilling}, J.} (\bibinfo{year}{1975}{\natexlab{a}}).
\newblock \bibinfo{title}{{Cosmic ray streaming. I - Effect of Alfven waves on
  particles}}.
\newblock {\it \bibinfo{journal}{\mnras}\/},  {\it \bibinfo{volume}{172}\/},
  \bibinfo{pages}{557--566}.

\bibitem[{{Skilling}(1975{\natexlab{b}})}]{Skill75c}
\bibinfo{author}{{Skilling}, J.} (\bibinfo{year}{1975}{\natexlab{b}}).
\newblock \bibinfo{title}{{Cosmic ray streaming. III - Self-consistent
  solutions}}.
\newblock {\it \bibinfo{journal}{\mnras}\/},  {\it \bibinfo{volume}{173}\/},
  \bibinfo{pages}{255--269}.

\bibitem[{Swaczyna et~al.(2022)Swaczyna, Schwadron, M{\"o}bius, Bzowski,
  Frisch, Linsky, McComas, Rahmanifard, Redfield, Winslow
  et~al.}]{Swaczyna2022}
\bibinfo{author}{Swaczyna, P.}, \bibinfo{author}{Schwadron, N.~A.},
  \bibinfo{author}{M{\"o}bius, E.} et~al. (\bibinfo{year}{2022}).
\newblock \bibinfo{title}{Mixing interstellar clouds surrounding the {Sun}}.
\newblock {\it \bibinfo{journal}{\apjl}\/},  {\it
  \bibinfo{volume}{937}\/}\bibinfo{issue}{(2)}, \bibinfo{pages}{L32}.

\bibitem[{{Vladimirov} et~al.(2012){Vladimirov}, {J{\'o}hannesson},
  {Moskalenko} \& {Porter}}]{VladimirMoskPamela11}
\bibinfo{author}{{Vladimirov}, A.~E.}, \bibinfo{author}{{J{\'o}hannesson}, G.},
  \bibinfo{author}{{Moskalenko}, I.~V.} et~al. (\bibinfo{year}{2012}).
\newblock \bibinfo{title}{Testing the origin of high-energy cosmic rays}.
\newblock {\it \bibinfo{journal}{\apj}\/},  {\it \bibinfo{volume}{752}\/},
  \bibinfo{pages}{68}. \DOIprefix\doi{10.1088/0004-637X/752/1/68}.
  \href{http://arxiv.org/abs/1108.1023}{\tt arXiv:1108.1023}.

\bibitem[{Wood et~al.(2002)Wood, M{\"u}ller, Zank \& Linsky}]{Wood2002}
\bibinfo{author}{Wood, B.~E.}, \bibinfo{author}{M{\"u}ller, H.-R.},
  \bibinfo{author}{Zank, G.~P.} et~al. (\bibinfo{year}{2002}).
\newblock \bibinfo{title}{Measured mass-loss rates of solar-like stars as a
  function of age and activity}.
\newblock {\it \bibinfo{journal}{\apj}\/},  {\it
  \bibinfo{volume}{574}\/}\bibinfo{issue}{(1)}, \bibinfo{pages}{412}.

\bibitem[{{Yoon} et~al.(2017){Yoon}, {Anderson}, {Barrau}, {Conklin}, {Coutu},
  {Derome}, {Han}, {Jeon}, {Kim}, {Kim}, {Lee}, {Lee}, {Lee}, {Lee}, {Link},
  {Menchaca-Rocha}, {Mitchell}, {Mognet}, {Nutter}, {Park}, {Picot-Clemente},
  {Putze}, {Seo}, {Smith} \& {Wu}}]{2017ApJ...839....5Y}
\bibinfo{author}{{Yoon}, Y.~S.}, \bibinfo{author}{{Anderson}, T.},
  \bibinfo{author}{{Barrau}, A.} et~al. (\bibinfo{year}{2017}).
\newblock \bibinfo{title}{Proton and helium spectra from the {CREAM-III}
  flight}.
\newblock {\it \bibinfo{journal}{\apj}\/},  {\it
  \bibinfo{volume}{839}\/}\bibinfo{issue}{(1)}, \bibinfo{pages}{5}.
  \DOIprefix\doi{10.3847/1538-4357/aa68e4}.
  \href{http://arxiv.org/abs/1704.02512}{\tt arXiv:1704.02512}.

\bibitem[{Zhang et~al.(2023)Zhang, Huang, Xu, Zhao \& Yuan}]{zhang2023galactic}
\bibinfo{author}{Zhang, R.}, \bibinfo{author}{Huang, X.}, \bibinfo{author}{Xu,
  Z.-H.} et~al. (\bibinfo{year}{2023}).
\newblock \bibinfo{title}{{Galactic} diffuse {\ensuremath{\gamma}}-ray emission
  from {GeV} to {PeV} energies in light of up-to-date cosmic-ray measurements}.
\newblock \href{http://arxiv.org/abs/2305.06948}{\tt arXiv:2305.06948}.

\bibitem[{Zhao et~al.(2022)Zhao, Liu, Yuan, Hu, Bi, Wu, Zhou \&
  Guo}]{Zhao_2022}
\bibinfo{author}{Zhao, B.}, \bibinfo{author}{Liu, W.}, \bibinfo{author}{Yuan,
  Q.} et~al. (\bibinfo{year}{2022}).
\newblock \bibinfo{title}{Geminga snr: Possible candidate of the local
  cosmic-ray factory}.
\newblock {\it \bibinfo{journal}{The Astrophysical Journal}\/},  {\it
  \bibinfo{volume}{926}\/}\bibinfo{issue}{(1)}, \bibinfo{pages}{41}. \URLprefix
  \url{https://dx.doi.org/10.3847/1538-4357/ac4416}.
  \DOIprefix\doi{10.3847/1538-4357/ac4416}.

\end{thebibliography}

\end{document}